%% file: paper.tex
\documentclass[11pt,oneside,reqno]{amsart}

\usepackage{adjustbox, algorithm, algpseudocode, amsfonts, amssymb, amsthm, amsmath, array, authblk, bm, booktabs, braket, caption, changepage, comment, dsfont, enumitem, etoolbox, fullpage, graphicx, mathrsfs, mathtools, microtype, multicol, multirow, pifont, placeins, pythonhighlight, relsize, subfloat, textcomp, url, xcolor, xpatch}

\usepackage{tikz, tikz-cd}
\usetikzlibrary{arrows.meta, positioning, shapes.geometric, decorations.pathreplacing}
\tikzset{
obs/.style={circle, draw, minimum size=8mm, inner sep=0pt},
lat/.style={circle, draw, dashed, minimum size=8mm, inner sep=0pt},
cd/.style={fill=gray!25},
resp/.style={circle, draw, double, minimum size=8mm, inner sep=0pt},
edge/.style={-{Stealth[length=2mm]}, thick}}

\usepackage{subcaption}

\usepackage[round]{natbib}
\usepackage{multibib}
\newcites{S}{References}

\setlist[enumerate]{leftmargin=1.5em, itemsep=0.5em}
\setlist[itemize]{leftmargin=1.5em, itemsep=0.5em}

\usepackage[foot]{amsaddr}

\usepackage{titlesec}

\titleformat{\section}
	{\titlerule\vspace{-2ex}}
	{\thesection.}{1ex}{\centering\scshape}
	[\vspace{.5ex}\titlerule]

\titleformat{\subsection}[runin]
  {\normalfont\normalsize\bfseries}{\thesubsection.}{1ex}{}[.]
\titlespacing*{\subsection}{0pt}{0.0\baselineskip}{0.5ex}

\titleformat{\subsubsection}[runin]
  {\normalfont\normalsize\itshape}{\thesubsubsection.}{1ex}{}
\titlespacing*{\subsection}{0pt}{0.0\baselineskip}{0.5ex}

\newtheorem{theorem}{Theorem}[section]

\newtheoremstyle{style}
  {\baselineskip} 
  {0em} 
  {\itshape} 
  {} 
  {\bfseries} 
  {.} 
  {.5em} 
  {} 
\theoremstyle{style}

\newtheorem*{theorem*}{Theorem}

\newtheorem*{definition*}{Definition}

\newtheorem*{lemma*}{Lemma}
\newtheorem{prop}[theorem]{Proposition}
\newtheorem*{prop*}{Proposition}

\usepackage{hyperref}
\hypersetup{
	colorlinks=true,
	linkcolor=black,
	citecolor=black,
	urlcolor=blue}

\usepackage[noabbrev,capitalise]{cleveref}
\crefname{prop}{Proposition}{Propositions}
\Crefname{prop}{Proposition}{Propositions}
\creflabelformat{equation}{#2\textup{#1}#3}

\algrenewcommand\algorithmicrequire{\textbf{Input:}}
\algrenewcommand\algorithmicensure{\textbf{Output:}}

\DeclareMathOperator{\1}{\mathds{1}}

\newcommand{\nindep}{\mkern3mu\not\mkern-3mu\perp\!\!\!\!\perp}

\newcommand{\yes}{\textcolor{green!55!black}{\ding{51}}}
\newcommand{\no}{\textcolor{red!75!black}{\ding{55}}}

\title{{\large P}rior-data fitted networks for variable selection}

\author{{\large O}mar {\large M}elikechi$^1$ \\
\rule{0pt}{2em}Department of Statistical Science, Duke University}
\address{$^1$omar.melikechi@duke.edu}

\begin{document}

\frenchspacing

\begin{abstract}
Prior-data fitted networks (PFNs) have become state-of-the-art for tabular prediction but remain underexplored in other areas of statistics. In this work, we present two PFNs for Bayesian variable selection: CD-PFN identifies conditional dependencies, and PC-PFN identifies parents and children in a causal graph. Both networks estimate posterior inclusion probabilities (PIPs) without posterior samples or data-specific modeling. In semi-synthetic and real data experiments, CD-PFN and PC-PFN often outperform existing methods, demonstrating the value of PFNs for variable selection and their potential in statistics more broadly. Open-source software is available.
\end{abstract}

\maketitle


\section{Introduction}\label{sec:introduction}


In Bayesian variable selection, the posterior inclusion probability (PIP) of a covariate quantifies its relevance to a response given observed data, providing an interpretable, variable-specific measure of uncertainty. Computing PIPs, however, typically requires specifying a likelihood and performing posterior inference using methods such as Markov chain Monte Carlo on each new dataset. This can be slow, and tractable inference often requires restrictive modeling assumptions.

Prior-data fitted networks (PFNs) offer an alternative approach to Bayesian inference in which a transformer is trained on data simulated from a prior to approximate the posterior \citep{muller2022}. Once trained, PFNs perform likelihood-free inference on new datasets in a single forward pass without posterior samples. Because training only requires sampling from the prior, PFNs readily accommodate a wide range of dataset characteristics that can complicate conventional posterior inference, such as nonlinearities, complex covariate and latent structures, and mixed data types. 

Currently, much of the PFN literature focuses on prediction, where models such as TabPFN, TabICL, and TabFM have achieved state-of-the-art performance on tabular data \citep{hollmann2025,qu2025,kong2026,erickson2025}. In this work, we instead study PFNs for variable selection, presenting two models for PIP estimation. The first, conditional dependence PFN (CD-PFN), estimates the posterior probability that a response is conditionally dependent on a covariate given all others. The second, parent--child PFN (PC-PFN), estimates the posterior probabilities that each covariate is a parent or child of the response in a directed acyclic graph.


\subsection*{Preview of results}


To preview the method, we present a semi-synthetic simulation study using gene expression data from The Cancer Genome Atlas (TCGA; \citealp{weinstein2013}). In each of 50 trials, $n=100$ samples and $p=50$ genes are randomly drawn from TCGA data, with five genes randomly chosen to determine the response. We consider three settings: one in which the response is a linear function of the five covariates, and two in which the relations are nonlinear. The first nonlinear setting uses functions included in the CD-PFN prior; the second uses held-out nonlinearities that CD-PFN did not train on. We compare CD-PFN to three other Bayesian methods---linear spike-and-slab \citep{mitchell1988}, Bayesian kernel machine regression (BKMR; \citealp{bobb2015}), and DART \citep{linero2018}---and to permutation importance rankings from TabPFN \citep{hollmann2025}. Full simulation details are in \cref{sec:illustrative-simulation}.

Without any tuning or dataset-specific modeling, CD-PFN nearly matches spike-and-slab on linear responses and performs best in both nonlinear settings (\cref{tab:blanket-tcga-illustration}). Its advantage over TabPFN in particular suggests that PFNs benefit from task-specific training, targeting selection rather than prediction in this case. Notably, CD-PFN is trained only on simulated data and thus has never seen gene expression measurements. Once trained, CD-PFN computes all 50 PIPs in 0.2 seconds on a CPU. TabPFN with permutation importance is over a thousand times slower, not because TabPFN is slow but because permutation importance requires over a thousand predictions per dataset.

\begin{table}[htbp]
\centering
\begin{adjustbox}{max width=\textwidth}
\begin{tabular}{lcccc}
\toprule
Method & Linear & Nonlinear (in-prior) & Nonlinear (held-out) & Runtime \\
\midrule
CD-PFN & $0.93\ [0.90,0.95]$ & $\mathbf{0.89}\ [0.87,0.92]$ & $\mathbf{0.58}\ [0.54,0.63]$ & $0.2/0.2/0.2$ \\
Spike-and-slab & $\mathbf{0.95}\ [0.93,0.97]$ & $0.35\ [0.30,0.40]$ & $0.22\ [0.19,0.25]$ & $0.2/0.1/0.1$ \\
BKMR & $0.90\ [0.86,0.93]$ & $0.72\ [0.68,0.76]$ & $0.41\ [0.37,0.45]$ & $6.2/6.2/6.2$ \\
DART & $0.60\ [0.54,0.66]$ & $0.47\ [0.40,0.54]$ & $0.33\ [0.29,0.38]$ & $3.9/3.8/3.7$ \\
TabPFN & $0.87\ [0.83,0.90]$ & $0.68\ [0.64,0.72]$ & $0.55\ [0.50,0.59]$ & $243/240/239$ \\
\bottomrule
\end{tabular}
\end{adjustbox}
\caption{\textit{Preview of results}. Average area under the precision--recall curve (AUPRC) over 50 trials for the linear, in-prior nonlinear, and held-out nonlinear settings, along with 95\% bootstrap confidence intervals. Higher AUPRC is better; bold marks the best AUPRC in each setting. CPU runtimes are reported in seconds per dataset and shown in the order of the simulation settings: linear/nonlinear-in-prior/nonlinear-held-out. CD-PFN, spike-and-slab, and BKMR rank covariates by their PIPs, DART by its inclusion probabilities, and TabPFN by permutation importance.}
\label{tab:blanket-tcga-illustration}
\end{table}


\subsection*{Related work}


Transformers trained on simulated data have been used for causal graph estimation (AVICI, \citealp{lorch2022}; DCD-PFN, \citealp{guan2026}) and conditional independence testing (ACID; \citealp{duong2025}). In concurrent work, LinearPFN \citep{schiekiera2026} computes PIPs for linear models with pairwise interactions, and SelectPFN \citep{luo2026} computes PIPs for sparse nonlinear models with continuous responses. Both of their priors generate responses directly from covariates and noise. By contrast, our priors allow covariates to be effects of the response or share unobserved causes with it. We also include a wide range of nonlinear effects, sparsity levels, covariate types, and both continuous and binary responses. At the time of writing, there is no publicly available software for DCD-PFN, SelectPFN, or ACID, so they are omitted from our empirical studies. \cref{tab:related-methods} provides an overview of different PFN selection methods.


\subsection*{Organization}


In \cref{sec:pfn}, we describe PFNs for binary parameters in general and for Bayesian variable selection specifically. In \cref{sec:model}, we describe the CD-PFN and PC-PFN priors. In \cref{sec:simulations,sec:real}, we report simulation and real data results. \cref{sec:conclusion} concludes.


\section{PFNs for variable selection}\label{sec:pfn}


We discuss PFNs for binary parameters in general (\cref{sec:binary}), and for CD-PFN and PC-PFN specifically (\cref{sec:cd,sec:pc}). We then cover PIP-based selection rules (\cref{sec:bvs}), the network architecture and training (\cref{sec:architecture-main}), and check the accuracy of CD-PFN's PIPs against an exact posterior while also demonstrating its fine-tuning abilities (\cref{sec:posterior-accuracy}).


\subsection{PFNs for binary parameters}\label{sec:binary}


Given an unknown binary parameter $\gamma\in\{0,1\}$ and observed data $D$, our objective is to estimate the posterior probability $\pi(D)=\mathbb{P}(\gamma=1\mid D)$. The standard Bayesian approach is to specify a prior $\Pi$ on a family of models $\mathcal{M}$, where each $M\in\mathcal{M}$ is a data-generating mechanism that determines $\gamma=\gamma(M)$. Integrating over models gives
\begin{align}\label{eq:integral}
\pi(D) &= \mathbb{P}(\gamma=1\mid D)
  = \int_{\mathcal{M}}\gamma(M)\,\Pi(dM\mid D),
\end{align}
where, by Bayes' rule, $\Pi(dM\mid D)\propto p(D\mid M)\,\Pi(dM)$ with $p(D\mid M)$ the likelihood of $D$ under $M$. The integral in \cref{eq:integral} is often intractable and must be approximated using methods such as Markov chain Monte Carlo (MCMC) or variational inference. Following \cite{muller2022}, we bypass this integral entirely by directly learning the map $D\mapsto\pi(D)$ as follows.

Since $\gamma$ is binary, its posterior distribution is Bernoulli with probability of success $\pi(D)$. To approximate this distribution, we train a model $\hat{\pi}_\phi$ with parameters $\phi$ that takes as input an arbitrary dataset $D$ and returns an estimate $\hat{\pi}_\phi(D)\in[0,1]$ of $\pi(D)$. One way to accomplish this is to minimize the expected Kullback--Leibler divergence
\begin{align}\label{eq:kl}
J(\hat{\pi}_\phi)=\mathbb{E}_D\!\left[\mathrm{KL}\!\left(\operatorname{Bern}(\pi(D))\,\|\,\operatorname{Bern}(\hat{\pi}_\phi(D))\right)\right]
\end{align}
between the true posterior and the network's estimate. Here, $\mathbb{E}_D$ denotes expectation over
\begin{align*}
p(D)=\int_{\mathcal{M}}p(D\mid M)\,\Pi(dM),
\end{align*}
the marginal distribution of datasets obtained by drawing $M\sim\Pi$ followed by $D\mid M\sim p(\cdot\mid M)$. This average is taken to accommodate a wide range of data rather than a single dataset.

The complicating factor in \cref{eq:kl} is that $J(\hat{\pi}_\phi)$ involves the unknown $\pi(D)$. However, this quantity can still be minimized by considering the binary cross-entropy loss
\begin{align*}
\ell(\gamma,q)=-\gamma\log q-(1-\gamma)\log(1-q),
\end{align*}
where $q\in[0,1]$ and $0\log 0=0$ and $-\log 0=\infty$. The following form of the PFN loss decomposition makes this connection explicit \citep{muller2022,gneiting2007}.

\begin{prop}\label[prop]{prop:pip-optimum}
For any function $q$ from datasets to $[0,1]$,
\begin{align*}
J(q)
&=\mathbb{E}[\ell(\gamma,q(D))]-\mathbb{E}_D[h(\pi(D))],
\end{align*}
where $h(u)=-u\log u-(1-u)\log(1-u)$ is binary entropy and $\mathbb{E}$ averages over the joint prior-data distribution of $(\gamma,D)$. Furthermore, $J(q)$ is minimized exactly when $q=\pi$ almost surely under $p(D)$.
\end{prop}

\begin{proof}
Expanding $J(q)$ gives
\begin{align*}
J(q)
  &=\int\left[\pi(D)\log\frac{\pi(D)}{q(D)}+(1-\pi(D))\log\frac{1-\pi(D)}{1-q(D)}\right]p(D)\,dD\\
  &=\int\left[-\pi(D)\log q(D)-(1-\pi(D))\log(1-q(D))-h(\pi(D))\right]p(D)\,dD\\
  &=\mathbb{E}[\ell(\gamma,q(D))]-\mathbb{E}_D[h(\pi(D))],
\end{align*}
where the last line holds because, by the law of total expectation, $\mathbb{E}_D[\ell(\pi(D),q(D))]=\mathbb{E}[\ell(\gamma,q(D))]$. Finally, since KL divergence is nonnegative and equals zero exactly when its two distributions agree, $J(q)=0$ exactly when $q(D)=\pi(D)$ for almost every $D$.
\end{proof}

\cref{prop:pip-optimum} justifies using $L(\phi)=\mathbb{E}[\ell(\gamma,\hat{\pi}_\phi(D))]$ as our training objective. \cref{alg:pfn-training} provides the full PFN training algorithm.

\begin{algorithm}[htbp]
\caption{(PFN training)}\label{alg:pfn-training}
\begin{algorithmic}[1]
\Require Prior $\Pi$, network $\hat{\pi}_\phi$, batch size $b$, number of updates $t_{\max}$
\State Initialize network parameters $\phi$
\For{$t=1,\dots,t_{\max}$}
    \State Draw models $M_1,\dots,M_b$ independently from $\Pi$
    \State Generate $D_k\sim p(\cdot\mid M_k)$ and set $\gamma_k=\gamma(M_k)$, for $k=1,\dots,b$
    \State Update $\phi$ using the gradient of $\frac{1}{b}\sum_{k=1}^b\ell(\gamma_k,\hat{\pi}_\phi(D_k))$
\EndFor
\Ensure Trained network $\hat{\pi}_\phi$
\end{algorithmic}
\end{algorithm}


\subsection{Conditional dependence}\label{sec:cd}


Let $X=(X_1,\dots,X_p)\in\mathcal{X}$ be covariates and $Y\in\mathcal{Y}$ a response. Write $[p]=\{1,\dots,p\}$ and, for $S\subseteq[p]$, let $X_S=\{X_j:j\in S\}$ and $X_{-j}=X_{[p]\setminus\{j\}}$. For a joint distribution $P$ on $\mathcal{X}\times\mathcal{Y}$, our first target indicates whether, under $P$, the covariate $X_j$ and the response $Y$ are conditionally dependent given the remaining covariates:
\begin{align*}
\gamma_j(P)=\1(Y\nindep X_j\mid X_{-j}).
\end{align*}
We place a prior $\Pi$ on a family $\mathcal{P}$ of joint distributions on $\mathcal{X}\times\mathcal{Y}$ and model $D=\{(x_i,y_i)\}_{i=1}^n$ as
\begin{align*}
P\sim\Pi \quad\text{and}\quad D\mid P\sim P^n.
\end{align*}
Here the generative model---denoted $M$ in \cref{sec:binary}---is the distribution $P$, and $\gamma_j(P)$ plays the role of $\gamma(M)$. The posterior inclusion probability (PIP) for $X_j$ is
\begin{align*}
\pi_j(D)=\mathbb{P}(\gamma_j(P)=1\mid D).
\end{align*}
Given new data, CD-PFN estimates one PIP per covariate in the dataset.


\subsection{Causal relations}\label{sec:pc}


Our second set of targets is to determine whether $X_j$ is a parent or child of $Y$ in a directed acyclic graph (DAG). Let $M$ be a structural model specifying a DAG $G_M$ and a corresponding distribution $P_M$ on $\mathcal{X}\times\mathcal{Y}$. Define
\begin{align*}
\gamma_j^{\mathrm{pa}}(M)&=\1(X_j\to Y\text{ in }G_M)
\quad\text{and}\quad
\gamma_j^{\mathrm{ch}}(M)=\1(Y\to X_j\text{ in }G_M).
\end{align*}
We place a prior $\Pi$ on a family $\mathcal{M}$ of such models and model the data $D$ as
\begin{align*}
M\sim\Pi \quad\text{and}\quad D\mid M\sim P_M^n.
\end{align*}
For $r\in\{\mathrm{pa},\mathrm{ch}\}$, the corresponding PIPs are
\begin{align*}
\pi_j^r(D)=\mathbb{P}(\gamma_j^r(M)=1\mid D).
\end{align*}
PC-PFN predicts parents and children, returning two estimated PIPs per covariate. Unlike the conditional dependence indicators, these are not determined by the joint distribution alone: models with different parent or child relations can induce the same $P_M$. Hence, the prior determines the relative posterior weights of such models when the data cannot distinguish them (\cref{sec:model}).


\subsection{False discovery control}\label{sec:bvs}


PIPs quantify the evidence for each covariate's relevance but do not by themselves offer a rule for selecting a set of covariates. One common approach is to select every covariate with PIP above 0.5. This gives the median probability model, which is optimal for prediction under certain conditions \citep{barbieri2004}. An alternative is to select covariates so as to control the false discovery rate (FDR)---the expected proportion of selected covariates that are null---which can be accomplished with PIPs as follows.

For any of the three targets, write $\gamma_j$ and $\pi_j$ for its inclusion indicator and PIP, and order the exact PIPs as $\pi_{j_{(1)}}\geq\cdots\geq\pi_{j_{(p)}}$. For a target FDR level $\alpha\in(0,1)$, let
\begin{align*}
k^*=\max\left\{k\in\{0,1,\dots,p\}:\sum_{i=1}^k(1-\pi_{j_{(i)}})\leq k\alpha\right\}.
\end{align*}
Then $S_\alpha=\{j_{(1)},\dots,j_{(k^*)}\}$ selects as many covariates as possible subject to an average posterior null probability of at most $\alpha$ \citep{newton2004}. For $S_\alpha\neq\emptyset$, the posterior expected false discovery proportion (FDP) satisfies
\begin{align*}
\mathbb{E}\big[\mathrm{FDP}(S_\alpha)\mid D\big]=\frac{1}{k^*}\sum_{i=1}^{k^*}(1-\pi_{j_{(i)}})\leq\alpha,
\end{align*}
where $\mathrm{FDP}(A)=\sum_{j\in A}(1-\gamma_j)/\max\{|A|,1\}$ for a set $A\subseteq[p]$. Averaging over the prior-data distribution gives Bayesian FDR at most $\alpha$, with no independence assumption on the $\gamma_j$. In practice, CD-PFN and PC-PFN use approximate PIPs, so this control is approximate.


\subsection{Architecture and training}\label{sec:architecture-main}


Any network that maps datasets to probabilities can, in principle, serve as $\hat{\pi}_\phi$. Ideally, $\hat{\pi}_\phi$ should accept datasets of arbitrary size and dimension, and each covariate's PIP should not depend on the order of the observations or covariates. We use transformer-based PFNs because they can be designed to meet these requirements and because of their success in prediction. Specifically, CD-PFN and PC-PFN use an architecture adapted from AVICI and TabPFN that alternates attention across observations and covariates \citep{lorch2022,hollmann2025}. Standardized values $x_{ij}$ and $y_i$ are embedded as tokens, and the response is distinguished from the covariates. After attention and pooling over observations, a shared output network combines each covariate's representation with the response representation to estimate its PIPs. CD-PFN and PC-PFN were trained on about 12 and 5 million simulated datasets, respectively, with the binary cross-entropy loss $\ell$ averaged over outputs, so that, by \cref{prop:pip-optimum}, each output targets its own PIP. Additional architecture and training details are in \cref{sec:architecture}.


\subsection{Posterior accuracy and fine-tuning}\label{sec:posterior-accuracy}


Computing exact PIPs is intractable under our priors, so our networks' PIPs cannot be checked against them directly. Instead, to assess posterior accuracy and the ability of PFNs to adapt to new priors, we fine-tune CD-PFN on a simpler prior whose exact PIPs can be computed, namely linear regression with Zellner's $g$-prior \citep{zellner1986,liang2008} and a beta-binomial prior on the set of active covariates \citep{scott2010}. The datasets have between $n=32$ and 512 observations and between $p=8$ and 12 correlated Gaussian covariates, so exact PIPs can be computed by enumerating all $2^p$ models (\cref{sec:posterior-accuracy-details}).

\begin{figure}[htbp]
\centering
\includegraphics[width=0.8\textwidth]{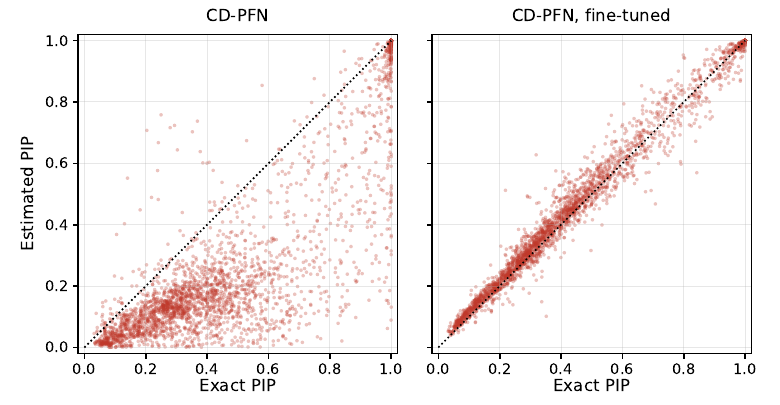}
\caption{\textit{Posterior accuracy}. CD-PFN's PIPs against exact PIPs under the linear $g$-prior for 3011 covariates across 300 datasets, before (left) and after (right) fine-tuning on that prior. The dotted line marks equality.}
\label{fig:posterior-accuracy}
\end{figure}

\cref{fig:posterior-accuracy} compares CD-PFN's PIPs with exact PIPs on 300 unseen datasets from this prior, before and after two hours of fine-tuning. Before fine-tuning, CD-PFN's PIPs reflect its own prior and are mostly too low, with a mean absolute error of 0.19. After fine-tuning, its PIPs lie much closer to the exact PIPs, with a mean absolute error of 0.03. Interestingly, most of the reduction in loss comes in the first minute of training (\cref{fig:ft-loss}). This example illustrates how users can adapt existing PFNs by training on a domain-specific prior, saving the cost of training from scratch.

\section{Priors}\label{sec:model}


Perhaps the most important component of a PFN is its prior, which generates the data upon which the network is trained. The prior encodes our beliefs about datasets the network will encounter in practice. In this work, to cover a wide range of datasets without committing to a single regression model or covariate structure, we combine two classes of data-generating mechanisms: an \textit{$X$-first class} that generates covariates before the response, and a class of \textit{structural causal models} (SCMs) that generate all variables jointly via a random DAG.

\begin{table}[htbp]
\centering
\small
\begin{tabular}{l>{\raggedright\arraybackslash}p{11.4cm}}
\toprule
Component & Specifications \\
\midrule
Response & Continuous (0.7) or binary (0.3) \\
\addlinespace
Functional forms & Linear (0.30), nonlinear additive (0.42), neural network (0.12), random Fourier features (0.08), products of functions of different inputs (0.08) \\
\addlinespace
Noise & Gaussian, Laplace, uniform, Student-$t$, or Gaussian mixture; $R^2$ (proportion of variance explained by parents) ranges from 0.05 to 0.95 \\
\addlinespace
Covariate types & Continuous, binary, or ordinal; in the $X$-first class, marginals may be skewed, heavy-tailed, or zero-inflated \\
\addlinespace
Covariate dependence & $X$-first: independent, autoregressive, factor, hierarchical factor, or nonlinear causal; SCM: random DAG with latent variables \\
\addlinespace
Covariates related to $Y$ & $X$-first: parents of $Y$, from none to half the covariates, with more weight on small sets; SCM: parents, children, and other parents of children, as given by the graph, plus covariates sharing a latent cause with $Y$ \\
\addlinespace
Sample size $n$ & 16 to 1024 \\
\addlinespace
Covariates $p$ & 8 to 256 (0.8) or 257 to 512 (0.2) \\
\addlinespace
Classes & CD-PFN: $X$-first (0.75), SCM (0.25); PC-PFN: $X$-first (0.425), SCM (0.425), linear Gaussian SCM (0.15) \\
\bottomrule
\end{tabular}
\caption{\textit{Summary of the prior}. Numbers in parentheses are prior probabilities; for example, 70\% of datasets drawn from the prior have a continuous response and 30\% have a binary response. Functional forms are drawn separately for each variable; for example, each variable is generated from its parents by a linear function with probability 0.3. Additional details about each component are in \cref{sec:prior-construction}.}
\label{tab:prior}
\end{table}

\cref{tab:prior} summarizes the CD-PFN and PC-PFN priors. In addition to covariates that cause the response, we include covariates that are effects of the response, such as a biomarker whose level is altered by a disease. We also include covariates that share an unobserved cause with the response, such as a biomarker driven by an unmeasured regulator that also affects the response. Modeling these relations explicitly in likelihood-based inference requires a joint distribution with latent variables and unknown causal directions, which can be difficult to specify and computationally challenging. A PFN, by contrast, only needs to simulate from such models.


\subsection{\texorpdfstring{$\bm{X}$}{X}-first}


The $X$-first class draws a joint distribution for $X$, chooses a subset $S\subseteq[p]$ of active covariates, and generates $Y=f(X_S,\varepsilon)$, where $\varepsilon$ is noise independent of $X$ and $f$ is drawn from the function library in \cref{tab:mechanisms}. By construction, $Y\nindep X_j\mid X_{-j}$ if and only if $j\in S$. The set $S$ is drawn independently of the distribution of $X$, so correlation among covariates is unrelated to the size of $S$. For PC-PFN, some covariates are also included as children of $Y$ (\cref{sec:xfirst-extensions}).


\subsection{Structural causal models}


The second class is inspired by TabPFN's SCM priors \citep{hollmann2025}. We draw a DAG over observed and latent variables and generate each variable from its parents. Graphs vary in density and include both roughly uniform connectivity and hubs. The response is a randomly chosen node, so it may have parents, children, or both. Latent nodes can be shared causes of a covariate and the response.

We allow for discrete variables in our SCMs, which appear either as truly discrete quantities or as coarsened continuous variables. The latter are common in practice, for example, when age is binned by decade. The distinction between truly discrete and coarsened matters because conditioning on a coarsened covariate may preserve dependencies that conditioning on its exact value would eliminate. Parent and child labels are read directly from the DAG, using edges between the underlying variables rather than their coarse measurements. Conditional dependence labels are also computed from the graph, accounting for latent variables and coarsening. \cref{fig:cd-graph} illustrates the ways a covariate can be conditionally dependent on the response, and \cref{sec:prior-labels} gives additional details.


\section{Simulation studies}\label{sec:simulations}


We evaluate CD-PFN and PC-PFN on semi-synthetic data generated from two real datasets. The first dataset consists of bulk RNA-seq gene expression measurements of 2000 genes in 889 kidney cancer patients from TCGA (\citealp{weinstein2013}). The second is the Bioresponse dataset from OpenML \citep{vanschoren2014}, which consists of 1391 measures of 3751 molecules' sizes, shapes, and compositions. These datasets pose distinct challenges: TCGA covariates are continuous and highly correlated, while most of the Bioresponse covariates are binary or integer-valued. 

Our simulation experiments cover a range of settings, including continuous and binary responses (regression and classification), and various sample sizes, numbers of covariates, and numbers of active covariates (\cref{tab:sim-settings}). Each experiment consists of 100 trials. In each trial, $n$ observations and $p$ covariates are randomly sampled from the real dataset and a synthetic response is generated from some of the sampled covariates using Gaussian processes. We evaluate CD-PFN on both datasets (\cref{sec:sim-cd}) and PC-PFN on the TCGA data and on a fully simulated DAG (\cref{sec:sim-pc}). 


\subsection{Methods and metrics}\label{sec:methods-metrics}


We compare CD-PFN to three Bayesian methods---linear spike-and-slab \citep{mitchell1988}, SuSiE \citep{wang2020}, and DART \citep{linero2018}---and to rankings from random forests \citep{breiman2001}, XGBoost \citep{chen2016}, lasso \citep{tibshirani1996}, partial and marginal correlations, ordinary least squares $t$-statistics, and mutual information. The FDR control analyses add three methods designed to control FDR: model-X knockoffs \citep{candes2018}, integrated path stability selection (IPSS; \citealp{melikechi2026,melikechi2025}), and the Benjamini--Hochberg procedure \citep{benjamini1995}. For parents and children, we compare PC-PFN to AVICI \citep{lorch2022}, Bayesian structure learning with BiDAG \citep{kuipers2022}, DAGMA \citep{bello2022}, the PC algorithm \citep{spirtes2000}, greedy equivalence search (GES; \citealp{chickering2002}), and max-min hill climbing (MMHC; \citealp{tsamardinos2006}). Software and specifications for all methods are listed in \cref{tab:methods,tab:methods-pc}. Two methods from \cref{sec:introduction}, TabPFN with permutation importance and BKMR, are excluded due to high computational costs. LinearPFN is also excluded, since it was trained on at most 30 covariates, fewer than in any of our settings.

Performance is measured by how well methods rank covariates and by selections made at target FDR levels. Specifically, ranking is assessed using the area under the precision--recall curve (AUPRC) where, at each ranking threshold, precision is defined as $\mathrm{TP}/(\mathrm{TP}+\mathrm{FP})$ and recall is $\mathrm{TP}/(\mathrm{TP}+\mathrm{FN})$, with TP, FP, and FN the numbers of true positives, false positives, and false negatives among the retained covariates. Varying the threshold traces out the precision--recall curve, and AUPRC is the area under it, with an AUPRC of one corresponding to perfect ranking. In the FDR control analyses, methods are evaluated by empirical FDR, the false discovery proportion of \cref{sec:bvs} averaged over trials, and the true positive rate (TPR), the recall averaged over trials.


\subsection{Conditional dependence}\label{sec:sim-cd}


In each trial, a random subset $S$ of the sampled covariates is chosen, and the response is generated from
\begin{align*}
f(X)=\sum_{j\in S}w_j\,g_j(X_j)+\sum_{(a,b)}w_{ab}\,h_{ab}(X_a,X_b),
\end{align*}
where each $g_j$ is a random function drawn from a Gaussian process, each $h_{ab}$ is a random function of two covariates drawn from a two-dimensional Gaussian process, the second sum is over two disjoint pairs of covariates in $S$, and the weights $w$ are random. Continuous responses are $f(X)$ plus Gaussian noise, and binary responses are Bernoulli with probability of success a logistic function of $f(X)$. Since $f$ depends only on $X_S$, $S$ is the conditional dependence set. \cref{fig:cd-gp-effects} shows examples of the functions $g_j$, and \cref{sec:cd-responses} gives the full construction. The baseline parameter values are $n=250$, $p=200$, and $\lvert S\rvert=10$; each is varied while keeping the other values fixed at baseline, giving a total of 30 experimental settings across the two datasets (TCGA or Bioresponse) and response types (regression or classification). Full details are in \cref{sec:cd-simulations}.

\begin{figure}[htbp]
\centering
\includegraphics[width=\textwidth]{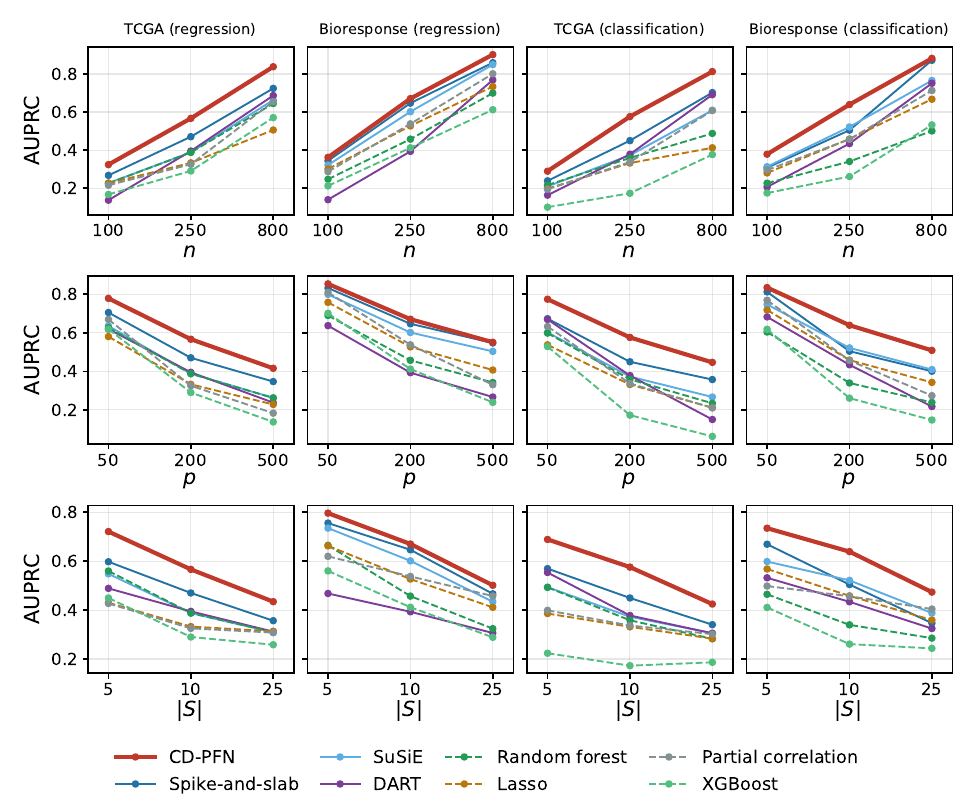}
\caption{\textit{Ranking accuracy}. Mean AUPRC over 100 trials as $n$, $p$, and $\lvert S\rvert$ vary from the baseline ($n=250$, $p=200$, $\lvert S\rvert=10$), for each dataset and response type.}
\label{fig:cd-auprc}
\end{figure}

\textit{Ranking}. CD-PFN has the highest AUPRC in all 30 settings, exceeding the next best method by more than two standard errors in 28 (\cref{fig:cd-auprc,tab:cd-paired,tab:cd-auprc-reg-tcga,tab:cd-auprc-reg-bio,tab:cd-auprc-cls-tcga,tab:cd-auprc-cls-bio}). This advantage persists at $p=1000$, which is nearly twice the largest dimension CD-PFN was trained on (\cref{tab:cd-auprc-reg-tcga,tab:cd-auprc-cls-tcga}). The closest competitor, spike-and-slab, nearly matches CD-PFN only for Bioresponse regression, whose many binary covariates render most effects linear. The TCGA baseline is also repeated with two other function families (\cref{tab:cd-families}). With purely linear effects, spike-and-slab ranks best for continuous responses and ties CD-PFN for binary ones, while CD-PFN is best in all other cases.

\begin{figure}[htbp]
\centering
\includegraphics[width=\textwidth]{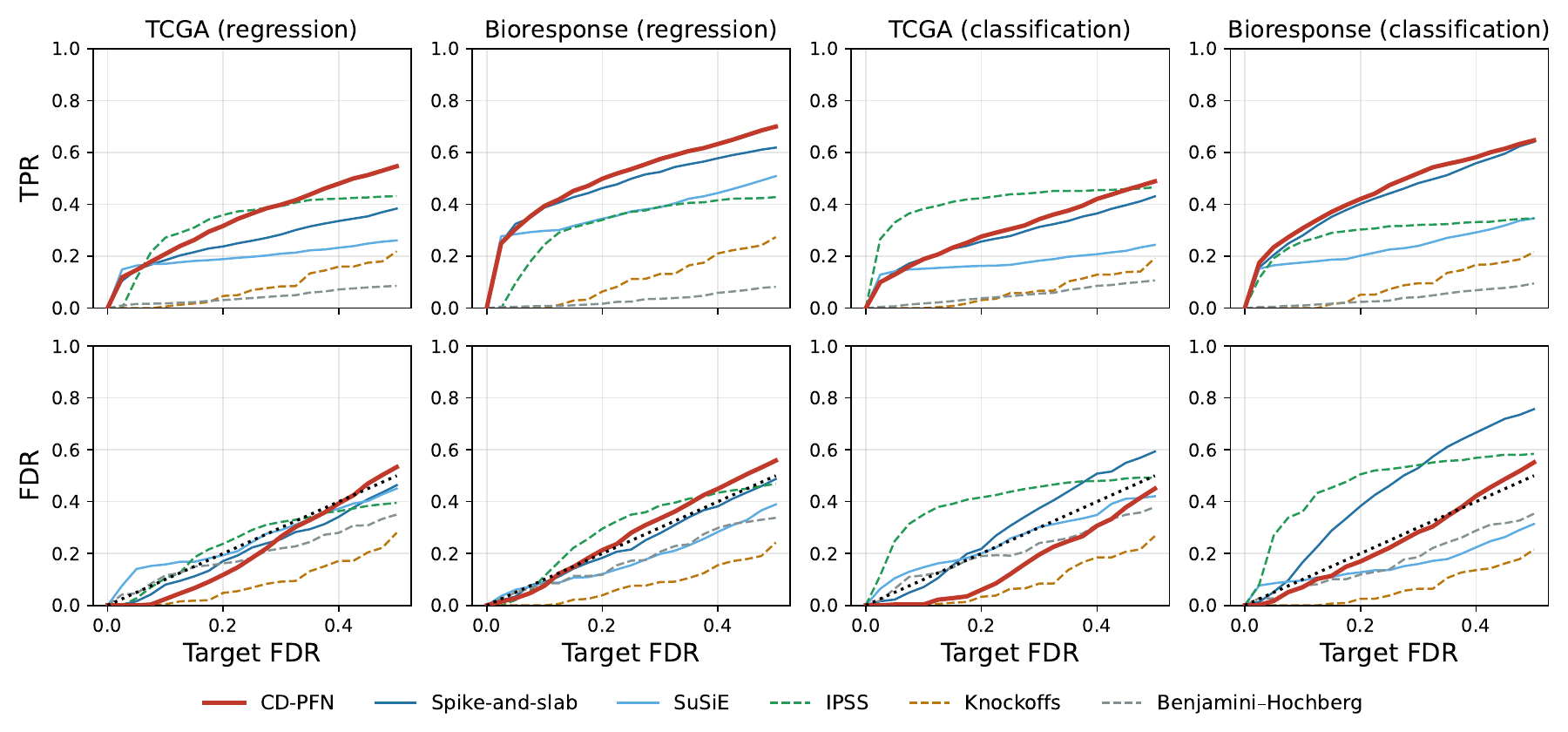}
\caption{\textit{FDR-based selection}. TPR (top) and empirical FDR (bottom) against the target FDR, averaged over 100 trials at the baseline simulation parameters. The dotted line marks an empirical FDR equal to the target; a method controls FDR if its empirical FDR curve lies at or below this line.}
\label{fig:cd-fdr}
\end{figure}

\textit{FDR control}. At target levels 0.05 and 0.1, CD-PFN's empirical FDR is below the target in all 30 settings. Among the methods whose empirical FDR is at most the target level of 0.1, CD-PFN has the highest TPR in 20 settings (\cref{fig:cd-fdr,tab:cd-fdr-reg-tcga,tab:cd-fdr-reg-bio,tab:cd-fdr-cls-tcga,tab:cd-fdr-cls-bio}). By contrast, at target level 0.1, spike-and-slab exceeds the target in 9 settings, SuSiE in 16, IPSS in 20, and Benjamini--Hochberg in 18. Two versions of knockoffs---one using lasso statistics, the other random forest importances---stay below the target but rarely select anything.

\subsection{Parents and children}\label{sec:sim-pc}


We evaluate PC-PFN on two designs. In the first, $n$ observations and $p$ covariates are sampled from the TCGA data, the response is generated from $\lvert\mathrm{pa}\rvert$ covariates as in \cref{sec:sim-cd}, and $\lvert\mathrm{ch}\rvert$ other covariates are made children of the response by adding to each a random function of the response drawn from a Gaussian process. The baseline parameter values are $n=250$, $p=50$, and three parents and three children ($\lvert\mathrm{pa}\rvert=3$ and $\lvert\mathrm{ch}\rvert=3$); each is varied while keeping the other values fixed at baseline, giving a total of nine settings. The second design is a simulated linear Gaussian DAG at the baseline values, which satisfies the assumptions of the classical causal discovery methods. Full details are in \cref{sec:pc-simulations}.

\textit{Ranking}. PC-PFN has the highest AUPRC for both parents and children in all nine TCGA settings, ranging from 0.67 to 0.95 for parents and from 0.95 to 1.00 for children (\cref{tab:pc-baseline,tab:pc-cells-auprc}). No other method exceeds 0.78 for parents or 0.90 for children in any of these settings. On the simulated linear Gaussian DAG, where BiDAG's model is correctly specified, BiDAG ranks best, with AUPRC 0.93 for parents and 0.96 for children, against 0.91 and 0.94 for PC-PFN (\cref{tab:pc-baseline}).

\textit{FDR control}. At target level 0.1, PC-PFN's empirical FDR is below the target for both parents and children in all ten settings, and among the methods whose empirical FDR is at most the target level, PC-PFN has the highest TPR in every TCGA setting (\cref{tab:pc-baseline,tab:pc-cells-parent,tab:pc-cells-child}). By contrast, BiDAG exceeds the target in every TCGA setting except the one without parents. DAGMA and the classical methods have no FDR target, and on the TCGA data their empirical FDR for parents reaches 0.89 and 0.74, respectively. AVICI stays below the target but rarely selects anything. In the simulated DAG experiment, PC-PFN ranks nearly as well as BiDAG but selects conservatively. Its empirical FDR is well below the target, and its TPR is lower than that of BiDAG for both parents and children. As in \cref{sec:posterior-accuracy}, this is consistent with PC-PFN's broad prior, which differs from the linear Gaussian model that BiDAG assumes.

\begin{table}[htbp]
\centering
\begin{adjustbox}{max width=\textwidth}
\input{tables/pc_baseline}
\end{adjustbox}
\caption{\textit{Parent and child results}. AUPRC and empirical FDR and TPR at target FDR level 0.1 at the TCGA baseline and on linear Gaussian DAGs. Bold marks the highest AUPRC and, among methods whose empirical FDR is at most 0.1, the highest TPR. PC, GES, and MMHC return graphs rather than scores, so they have no AUPRC; PC with kernel tests and GES were too slow for the TCGA design.}
\label{tab:pc-baseline}
\end{table}


\section{Real data}\label{sec:real}


We now study real responses whose relations to certain covariates are established, for example, by independent experiments or domain knowledge. In \cref{sec:real-cd}, we apply CD-PFN to 32 biomarker tasks from TCGA and cancer cell lines in DepMap and GDSC \citep{tsherniak2017,iorio2016}, and to a metabolic syndrome study from the National Health and Nutrition Examination Survey (NHANES; \citealp{nhanes2020}). In \cref{sec:real-pc}, we apply PC-PFN to relations of known direction in NHANES and to a physical system with a known causal graph. Details about each of these datasets and additional results are in \cref{sec:real-data}.


\subsection{Conditional dependence}\label{sec:real-cd}


Covariates in the 32 biomarker tasks are genes in the MSK-IMPACT cancer gene panel \citep{cheng2015}. Nine of the 32 tasks come from TCGA, where the response is a tumor status, covariates are 456 panel genes, and established biomarkers are genes that define or reflect tumor status. For example, the response in one such task is estrogen receptor status in breast cancer and the established covariate is ESR1, the gene that encodes this receptor \citep{green1986}. The other 23 tasks come from cancer cell line data, where the response is a cell line's dependence on a gene or its sensitivity to a drug, and the covariates are 504 panel genes and their mutations. There is exactly one established biomarker in 25 of the 32 tasks. The other seven tasks each have two or three established biomarkers, for a total of 18 altogether. 

The NHANES data have $n=2076$ participants and $p=65$ variables consisting of age, sex, ethnicity, education, and 61 exam and laboratory measurements. The response is whether a participant has metabolic syndrome, which is defined by thresholds on waist circumference, triglycerides, HDL cholesterol, systolic and diastolic blood pressure, and fasting glucose, with waist and HDL thresholds depending on sex \citep{alberti2009}. These seven defining variables, including sex, are therefore relevant to metabolic syndrome by construction. To increase the robustness of our analysis, we apply each selection method to 20 random samples of 1000 participants, rather than to the full dataset.

Results for the biomarker and metabolic syndrome studies are reported in \cref{tab:rd-summary}. Since false discoveries are difficult to verify in these data, we assess performance by the number of established covariates selected, with the number of other selections reported as a measure of parsimony rather than error. CD-PFN selects the established biomarker in 21 of 25 single-biomarker tasks; spike-and-slab, SuSiE, and IPSS select it in 22 or 23, with two to four times as many other selections on average. Among tasks with several established biomarkers, CD-PFN selects 15 of the 18, against at most 13 for the other methods. For metabolic syndrome, CD-PFN selects 5.8 of the seven defining variables on average over the 20 random samples, against 4.5 for spike-and-slab and 4.3 for SuSiE; IPSS selects 6.0, with twice as many other selections.

\begin{table}[htbp]
\centering
\begin{adjustbox}{max width=\textwidth}
\input{tables/rd_summary}
\end{adjustbox}
\caption{\textit{Selection results}. Established covariates selected at target FDR level 0.1, with the average number of additional selections in parentheses. The biomarker columns give totals over the 25 tasks with one established biomarker and the 7 tasks with several. The metabolic syndrome column reports the average number of established covariates selected over 20 random samples of 1000 NHANES participants.}
\label{tab:rd-summary}
\end{table}


\subsection{Parents and children}\label{sec:real-pc}


In NHANES, no measurement can cause age, sex, or ethnicity. We use this fact to conduct two experiments on all 2076 participants, about twice the largest sample size in PC-PFN's training prior. In the first, each of the 61 measurements serves in turn as the response, with the other 64 variables as covariates, so age, sex, and ethnicity can be parents of the response but not children. In the second, each of age, sex, and ethnicity serves in turn as the response, with the other 64 variables as covariates, so the measurements can be children but not parents.

\begin{figure}[htbp]
\centering
\includegraphics[width=\textwidth]{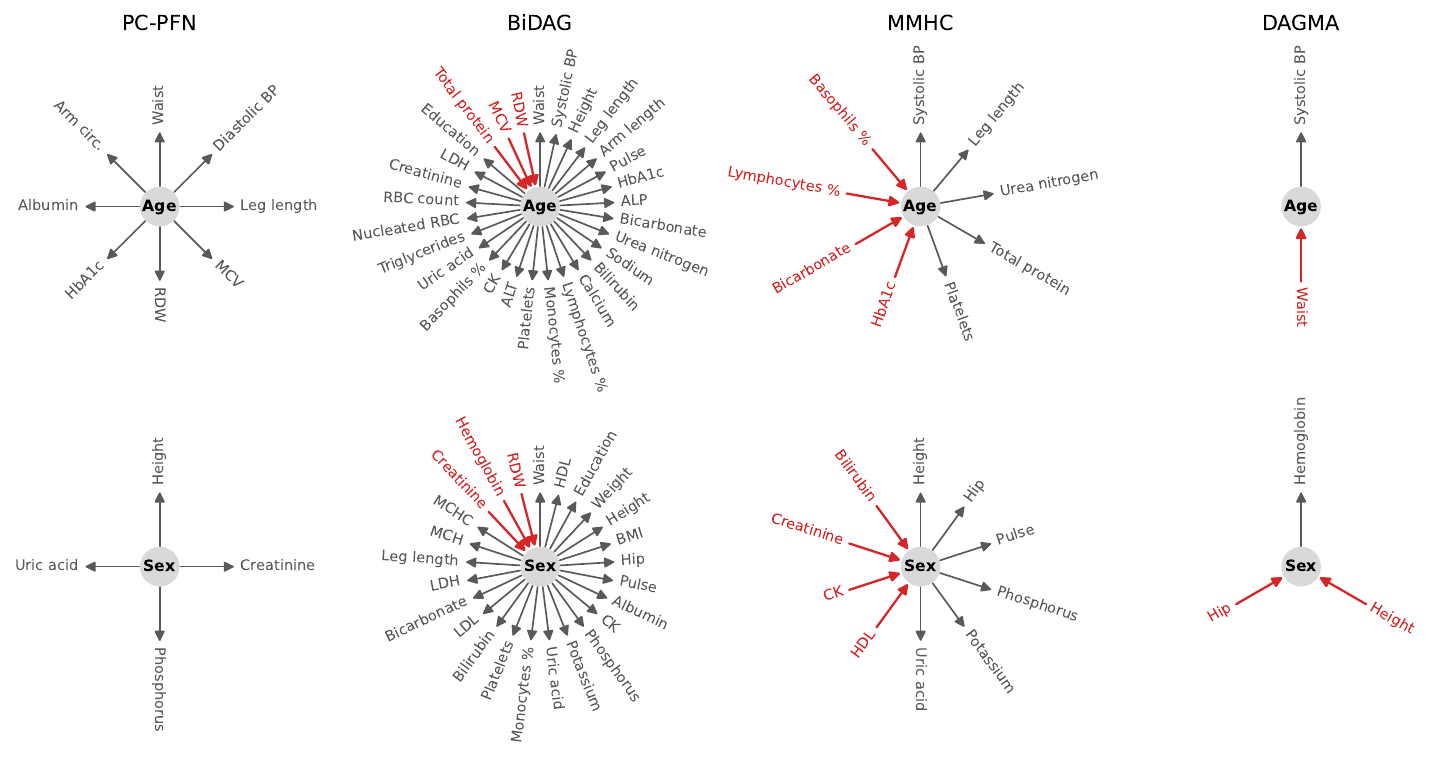}
\caption{\textit{NHANES selections with age and sex as the response}. Covariates selected at target FDR level 0.1 with age (top) or sex (bottom) as the response. Arrows out of the response are children and arrows into it are parents. Since no measurement can cause age or sex, red arrows are errors. AVICI, which selects nothing, is omitted.}
\label{fig:rd-exogenous}
\end{figure}

All 45 of PC-PFN's selections involving age, sex, or ethnicity have the correct direction (\cref{tab:rd-exogenous}). DAGMA and MMHC orient these relations correctly only about half the time, and AVICI makes no such selections. BiDAG took between 1 and 2.5 hours per dataset (PC-PFN took under 15 seconds), so it was run only with age, sex, and ethnicity as responses, where 8 of its 58 selections had the wrong direction. \cref{fig:rd-exogenous} shows the selections with age and sex as the response. Those of PC-PFN are plausible, such as effects of sex on height, creatinine, and uric acid, and of age on blood pressure and HbA1c. By contrast, each other method calls some measurements causes of age or sex.

In our second causal study, we investigate the light tunnel experiment of \citet{gamella2025}, consisting of a physical apparatus in which a light source shines through two polarizers onto light sensors. Its inputs, including the brightness of the light source and the angles of the polarizers, are set at random, and its causal graph of 18 variables and 37 edges is known by construction. Each variable serves in turn as the response, on all 10{,}000 observations, nearly ten times the largest sample size in the training prior.

PC-PFN ranks parents about as well as BiDAG and ranks children best (\cref{tab:rd-chamber}). PC-PFN and GES are the only methods that select parents with empirical FDR below the target, and GES finds more of them. No method selects children reliably; AVICI's empirical FDR for children is zero only because it selects no children at all, and every other method exceeds the target.

\begin{table}[htbp]
\centering
\begin{adjustbox}{max width=\textwidth}
\input{tables/rd_chambers_pc}
\end{adjustbox}
\caption{\textit{Light tunnel results}. AUPRC, and empirical FDR and TPR at target FDR level 0.1, with each of the 18 variables as the response on all 10{,}000 observations. FDR is averaged over all responses and TPR over those with parents or children. Bold marks the highest AUPRC and, among methods whose empirical FDR is at most 0.1, the highest TPR, including ties.}
\label{tab:rd-chamber}
\end{table}


\section{Discussion}\label{sec:conclusion}


We introduced two PFNs for Bayesian variable selection, CD-PFN and PC-PFN, and demonstrated their efficacy on semi-synthetic and real data. Two directions for future work stand out. The first is to better understand how the prior affects performance. A broad prior yields a general-purpose model, but substantive knowledge may justify a narrower class of data-generating mechanisms for a particular application. In our posterior accuracy experiment, fine-tuning CD-PFN on a linear regression prior substantially improved its accuracy under that prior. For prediction, \citet{regenwetter2026} similarly adapt TabPFN to engineering problems by training it on synthetic data that resemble engineering data. Future work should develop principled strategies for such adaptation, including how to construct application-specific priors, how much simulated data is required, and which parts of a pretrained model should be fine-tuned. Training on larger datasets is another natural extension, though our preliminary results show that CD-PFN and PC-PFN extrapolate well to sample sizes and dimensions that are larger than any they trained on.

The second direction follows from the general formulation in \cref{sec:binary}. While this work focused on conditional dependence, parents, and children, any binary parameter can serve as the inferential target in that framework. Even more broadly, our results contribute to growing evidence that PFNs are a promising new tool for Bayesian inference; understanding and developing them further is an interesting and important next step \citep{muller2025position}.


\section*{Code and data availability}


The \texttt{varselpfn} package and trained networks can be installed from PyPI (\texttt{pip install varselpfn}) and from \url{https://github.com/omelikechi/varselpfn}. Code to reproduce all results is at \url{https://github.com/omelikechi/varselpfn-paper}, and saved results are archived at \url{https://doi.org/10.5281/zenodo.23920743}. All datasets are publicly available from their original sources.

\bibliographystyle{plainnat}
\bibliography{refs}

\clearpage

\setcounter{page}{1}
\setcounter{section}{0}
\setcounter{equation}{0}
\setcounter{table}{0}
\setcounter{figure}{0}
\setcounter{algorithm}{0}
\renewcommand{\thepage}{S\arabic{page}}
\renewcommand{\thesection}{S\arabic{section}}
\renewcommand{\theequation}{S\arabic{equation}}
\renewcommand{\thetable}{S\arabic{table}}
\renewcommand{\thefigure}{S\arabic{figure}}
\renewcommand{\thealgorithm}{S\arabic{algorithm}}
\renewcommand{\theHsection}{S\arabic{section}}
\renewcommand{\theHsubsection}{S\arabic{section}.\arabic{subsection}}
\renewcommand{\theHequation}{S\arabic{equation}}
\renewcommand{\theHtable}{S\arabic{table}}
\renewcommand{\theHfigure}{S\arabic{figure}}
\renewcommand{\theHalgorithm}{S\arabic{algorithm}}
\renewcommand{\topfraction}{0.95}
\renewcommand{\bottomfraction}{0.95}
\renewcommand{\textfraction}{0.05}
\renewcommand{\floatpagefraction}{0.75}
\setcounter{topnumber}{5}
\setcounter{bottomnumber}{5}
\setcounter{totalnumber}{10}

\begin{center}
{\bfseries\Large{S}\bfseries\large{UPPLEMENTARY MATERIAL}}

\vspace{2em}



\end{center}



\section{Related methods}\label{sec:related-methods}


\cref{tab:related-methods} compares CD-PFN and PC-PFN with the amortized methods discussed in \cref{sec:introduction}.

\begin{table}[htbp]
\centering
\footnotesize
\begin{adjustbox}{max width=\textwidth}
\begin{tabular}{l>{\raggedright\arraybackslash}p{3.8cm}ccccccc}
\toprule
 & & Nonlinear & Children of & Binary & Causal & & & \\
Method & Target & effects & response & response & outputs & $n$ & $p$ & Code \\
\midrule
CD-PFN & Conditional dependence & \yes & \yes & \yes & \no & 16--1024 & 8--512 & \yes \\
PC-PFN & Parents and children & \yes & \yes & \yes & \yes & 16--1024 & 8--512 & \yes \\
\midrule
LinearPFN & Linear regression inclusion & \no & \no & \no & \no & 20--2000 & 2--30 & \yes \\
SelectPFN & Nonlinear regression inclusion & \yes & \no & \no & \no & 50--1000 & 5--500 & \no \\
DCD-PFN & Markov boundaries; causal graph & \yes & \yes & -- & \yes & -- & -- & \no \\
AVICI & Causal graph & \yes & \yes & \no & \yes & 200 & 2--50 & \yes \\
ACID & Test of one conditional independence & \yes & \yes & \no & \no & 50, 200 & 5--20 & \no \\
\bottomrule
\end{tabular}
\end{adjustbox}
\caption{\textit{Amortized selection methods}. Targets and training priors reported in each paper. All methods return probabilities except ACID, which returns $p$-values. ``Children of response" indicates whether the prior includes covariates caused by the response. A Markov boundary is a minimal set of variables given which the response is independent of all others \protect\citepS{pearl1988}. The columns $n$ and $p$ give training ranges for sample size and number of variables; for SelectPFN, $n$ counts both training and query rows, and for ACID, $p$ is the dimension of the conditioning set, with one-dimensional $X$ and $Y$. ``Code" indicates whether code or trained models are publicly available at the time of writing. Dashes indicate unreported values.}
\label{tab:related-methods}
\end{table}


\section{Preview simulation}\label{sec:illustrative-simulation}


We generate 50 datasets in each response setting, with $n=100$ observations and $p=50$ covariates. For each replicate, we draw a cancer type (bladder, glioma, or kidney), a measurement type (RNA-seq or miRNA), 100 patients, and 50 of the 500 most variable measurements without missing values. We standardize each covariate. The cancer type, measurement type, patients, and covariates are redrawn in every replicate. A set of active covariates $S\subseteq[p]$ of size five is drawn uniformly without replacement. Coefficient magnitudes are drawn independently from $\operatorname{Unif}(0.5,1.5)$ and their signs independently with equal probability. The linear signal is
\begin{align*}
f_{\mathrm{lin}}(X)=\sum_{j\in S}\beta_jX_{j}.
\end{align*}
In the in-prior nonlinear setting, each active covariate is assigned independently and uniformly to one of the functions
\begin{align*}
\mathcal{H}_{\mathrm{in}}=\left\{x^2,\ |x|,\ \exp(-x^2/2)\right\}.
\end{align*}
In the held-out nonlinear setting, each active covariate gets a randomly drawn natural cubic spline, kept only if it is poorly approximated by every function in the prior's library. Within each nonlinear setting, we standardize every transformed term and set
\begin{align*}
f(X)=\sum_{j\in S}\beta_jh_j(X_j).
\end{align*}
We add Gaussian noise with $\operatorname{Var}(f)/\operatorname{Var}(\varepsilon)=2$. The three settings share the same covariates, support, and coefficients.


\subsection{Methods}


CD-PFN is applied as pretrained, without fitting or tuning on the simulated datasets. Spike-and-slab regression (BAS; \citealpS{clyde2025}), DART \citepS{linero2018}, and BKMR \citepS{bobb2015} use the settings in \cref{tab:methods}, with $\operatorname{Beta}(1,9)$ priors on the spike-and-slab and BKMR inclusion rates, whose mean matches the proportion of active covariates. TabPFN \citepS{hollmann2025} ranks covariates by cross-fitted permutation importance and returns no probabilities, so only its AUPRC is reported. \cref{tab:blanket-tcga-illustration} reports mean AUPRC with 95\% bootstrap confidence intervals from 10{,}000 resamples of the 50 datasets in each setting. Runtimes are mean wall-clock times per dataset on the same Apple silicon CPU. TabPFN was timed on one thread for one dataset per setting.


\section{Architecture and training}\label{sec:architecture}


CD-PFN and PC-PFN share a transformer architecture that alternates attention across observations and covariates, as in TabPFN v2 and AVICI \citepS{hollmann2025,lorch2022}. Our implementation builds on nanoTabPFN \citepS{pfefferle2025}, and \cref{tab:architecture} lists its dimensions.

Each dataset is arranged as a matrix $z\in\mathbb{R}^{n\times(p+1)}$, with covariates in the first $p$ columns and the response in the last. Columns are standardized as $u_{ij}=(z_{ij}-\bar z_j)/s_j$, where $\bar z_j$ and $s_j$ are the column's sample mean and standard deviation. Rescaling a variable does not change its conditional dependence on the response or whether it is a parent or child of the response, so standardization leaves the targets unchanged. Each standardized entry is embedded as a token
\begin{align*}
h_{ij}^{(0)}=w u_{ij}+b+e_{\1(j=p+1)}\in\mathbb{R}^d,
\end{align*}
where $d=192$, $w,b\in\mathbb{R}^d$ are shared across entries, and the role embeddings $e_0,e_1\in\mathbb{R}^d$ distinguish the covariates from the response. There are no position embeddings. Each transformer block applies self-attention across covariates within each observation, self-attention across observations within each covariate, and a feedforward network, each with layer normalization \citepS{ba2016} and a residual connection. After the last block, each column is summarized by the mean and coordinatewise maximum of its tokens over observations, giving $\bar h_j\in\mathbb{R}^{2d}$. A two-layer network, shared across covariates, maps each covariate's summary $\bar h_j$ together with the response's $\bar h_{p+1}$ to three logits, for conditional dependence, parent, and child; their logistic transforms are the estimated PIPs. Because there are no position embeddings and the output network is shared, the network is invariant to the order of observations and equivariant to the order of covariates.

\begin{table}[htbp]
\centering
\begin{tabular}{ll}
\toprule
Component & Setting \\
\midrule
Token dimension $d$ & 192 \\
Transformer blocks & 8 \\
Attention heads & 6 \\
Feedforward hidden dimension & 768 \\
Trainable parameters & 5{,}341{,}827 \\
\bottomrule
\end{tabular}
\caption{\textit{Network dimensions}. Shared by CD-PFN and PC-PFN.}
\label{tab:architecture}
\end{table}

We trained each PFN on a single NVIDIA RTX A5000 or RTX 5000 Ada GPU using AdamW \citepS{loshchilov2019} with a warmup--stable--decay learning-rate schedule. Total training time was about 72 GPU hours for CD-PFN and 44 for PC-PFN. Each batch contained freshly simulated datasets with a common $n$ and $p$; across batches, $n$ ranged from 16 to 1024 and $p$ from 8 to 512. We selected the checkpoint with the lowest loss on held-out simulated datasets. The loss was binary cross-entropy averaged over covariates and, for PC-PFN, over the parent and child outputs.

Both models were initialized from a common network trained on all three outputs. CD-PFN was then trained on the conditional dependence output only, first for 24 hours on $X$-first draws without children of the response and then for 24 hours on a prior containing 75\% such draws and 25\% SCMs. PC-PFN was trained on the parent and child outputs only, using equal shares of the two classes and including children of the response in $X$-first draws (\cref{sec:xfirst-extensions}). It was then trained for 12 additional hours on a prior with more parents and children of the response and a linear Gaussian component. Configuration files for both models are provided with the code.


\section{Posterior accuracy}\label{sec:posterior-accuracy-details}

\begin{figure}[htbp]
\centering
\includegraphics[width=0.6\textwidth]{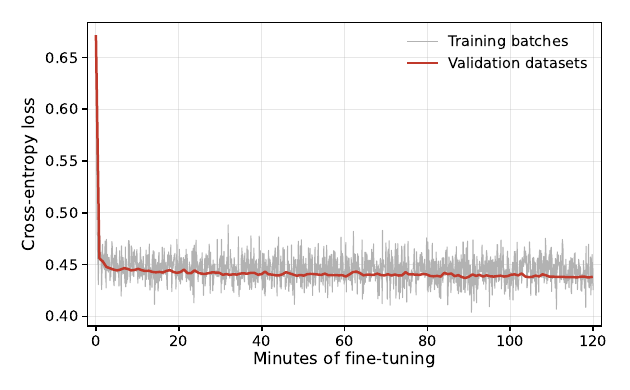}
\caption{\textit{Fine-tuning loss}. Cross-entropy loss of CD-PFN's conditional dependence output during fine-tuning on the linear $g$-prior, on training batches (gray) and on 512 fixed validation datasets (red).}
\label{fig:ft-loss}
\end{figure}

The prior of \cref{sec:posterior-accuracy} is linear regression with Zellner's $g$-prior \citepS{zellner1986}. The number of covariates is uniform on $\{8,9,10,11,12\}$, the sample size is log-uniform on $[32,512]$, and the covariates follow a Gaussian factor model with between zero and three factors. The response is $y=\alpha+X\beta+\sigma\varepsilon$ with $\varepsilon\sim\mathrm{N}(0,I_n)$, a beta-binomial$(1,1)$ prior on the active set $\gamma$, $\beta_\gamma\mid g,\sigma\sim\mathrm{N}(0,g\sigma^2(\tilde X_\gamma^\top\tilde X_\gamma)^{-1})$ for the centered covariates $\tilde X$, $g/n$ uniform on nine log-spaced values from 0.01 to 1, and uninformative priors $p(\alpha)\propto1$ and $p(\sigma^2)\propto1/\sigma^2$. Under the priors on $\alpha$ and $\sigma^2$, shifting or rescaling the response or any covariate leaves the posterior of $\gamma$ unchanged, so the posterior can be computed from the standardized data seen by the network. We obtain exact PIPs by enumerating all $2^p$ models and using the closed-form marginal likelihood given $g$ \citepS{liang2008}. Fine-tuning started from CD-PFN's weights and used loss on the conditional dependence output, which equals $\gamma$ under this prior. \cref{fig:ft-loss} shows this loss over the two hours of fine-tuning.

\section{Prior construction}\label{sec:prior-construction}



Across all classes, the sample size $n$ is drawn uniformly on $\{16,\dots,1024\}$ with probability 0.5 and log-uniformly otherwise. The number of covariates $p$ is drawn uniformly on $\{257,\dots,512\}$ with probability 0.2 and otherwise on $\{8,\dots,256\}$ using the same uniform/log-uniform mixture.


\subsection{Mechanisms and standardization}


Nonlinear additive mechanisms combine randomly scaled and shifted functions, including monotone, periodic, threshold, and localized effects. Multivariate mechanisms use small randomly initialized neural networks, random Fourier features, or products of functions of different inputs. Product mechanisms may also include additive effects from the remaining inputs. The piecewise-constant family sums random step functions of individual inputs. \cref{tab:mechanisms} lists every family with its prior weight and random parameters.

\begin{table}[htbp]
\centering
\small
\begin{tabular}{>{\raggedright\arraybackslash}p{2.6cm}c>{\raggedright\arraybackslash}p{5.4cm}>{\raggedright\arraybackslash}p{5.4cm}}
\toprule
Family & Weight & Mechanism $f(u_1,\ldots,u_d)$ & Random parameters \\
\midrule
Linear & 0.30 & $d^{-1/2}\sum_{i} w_i u_i$ & $|w_i|\sim\mathrm{U}(0.5,1.5)$, random sign \\
Additive & 0.30 & $d^{-1/2}\sum_{i} w_i\, g_i(s_i u_i+b_i)$ & $g_i$ drawn from the edge functions below; $w_i$ as for linear; $s_i$ log-uniform on $[0.5,2]$; $b_i\sim\mathrm{U}(-1,1)$, and $b_i=0$ for the identity \\
Neural network & 0.12 & $v^\top h/\sqrt{m}$, with $h=a(h'W+c)$ applied once or twice, starting from $h'=u$ & Width $m$ uniform on $\{8,\ldots,32\}$ per layer; $W$ with entries $\mathrm{N}(0,1/\text{input width})$; $c\sim\mathrm{N}(0,0.25\,I)$; $v\sim\mathrm{N}(0,I)$; $a$ one of tanh, ReLU, sigmoid, Gaussian bump, softplus, absolute value \\
Random Fourier features & 0.08 & $64^{-1/2}\sum_{l=1}^{64} v_l\cos(\omega_l^\top u+\phi_l)$ & $\omega_l\sim\mathrm{N}(0,\ell^{-2}I)$ with $\ell$ log-uniform on $[1,4]$; $\phi_l\sim\mathrm{U}(0,2\pi)$; $v_l\sim\mathrm{N}(0,1)$ \\
Piecewise constant & 0.12 & $\sum_{i} c_i(u_i)$, each $c_i$ a step function & 1 to 7 cut points from $\mathrm{U}(-1.5,1.5)$; level values $\mathrm{N}(0,1)$ \\
Product & 0.08 & $\sum_{t=1}^{T} g_t(u_{a_t}+b_t)\,h_t(u_{a'_t}+b'_t)$, plus an additive mechanism in the remaining inputs & Disjoint random pairs $(a_t,a'_t)$; $g_t,h_t$ drawn from identity, tanh, sigmoid, Gaussian bump, ReLU; $b_t,b'_t\sim\mathrm{U}(-1,1)$; $T=1$, except for the $X$-first response, where $T$ is uniform on $\{1,\ldots,\min(3,\lfloor\lvert S\rvert/2\rfloor)\}$; additive when $\lvert S\rvert=1$ \\
\midrule
\multicolumn{4}{l}{\textit{Edge functions $g$ of the additive family, with their probabilities}} \\
\midrule
\multicolumn{4}{>{\raggedright\arraybackslash}p{14.4cm}}{Identity $x$ (0.12); tanh (0.12); sigmoid $(1+e^{-x})^{-1}$ (0.12); Gaussian bump $e^{-x^2/2}$ (0.12); ReLU $\max(x,0)$ (0.10); softplus $\log(1+e^{x})$ (0.10); square $x^2$ (0.10); absolute value $|x|$ (0.10); step $\1(x>0)$ (0.08); sine (0.02); cosine (0.02).} \\
\bottomrule
\end{tabular}
\caption{\textit{Mechanisms relating a variable to its inputs}. The inputs $u_1,\ldots,u_d$ are the standardized values of the variable's parents (for the $X$-first response, the covariates in $S$). One family is drawn per variable with the stated weights, and its parameters are drawn independently for each variable and dataset. Every input is checked to remain functionally active, meaning that shuffling it changes the observed output after any discretization, with the mechanism redrawn otherwise. The output is standardized and combined with standardized noise as $\sqrt{1-r}\,f+\sqrt{r}\,\varepsilon$, with $r\sim\mathrm{U}(0.05,0.95)$ and $\varepsilon$ either Gaussian, Laplace, uniform, Student-$t$ with between 2.5 and 10 degrees of freedom, or a two-component Gaussian mixture. A binary response thresholds this value at a reference quantile drawn from $\mathrm{U}(0.2,0.8)$.}
\label{tab:mechanisms}
\end{table}

For each nonroot continuous node, the simulator standardizes the signal and noise separately and combines them as $\sqrt{1-r}\,f+\sqrt{r}\,\varepsilon$, with $r\in[0.05,0.95]$. The result is standardized before being passed to downstream mechanisms. Centers and scales are estimated from 512 reference observations generated alongside, but excluded from, the training dataset. Binary and ordinal values are obtained by thresholding at reference quantiles. When discretization is part of the mechanism, descendants receive the standardized discrete value; when it represents coarse measurement, descendants receive the underlying continuous value.


\subsection{\texorpdfstring{$\bm{X}$}{X}-first extensions}\label{sec:xfirst-extensions}


Factor designs vary the number of factors, loadings, and idiosyncratic noise. The hierarchical variant combines a few global factors with group-specific factors, producing both broad correlation and tighter clusters. The response support is drawn independently of the covariate distribution, with positive probability on the empty set and a mixture of uniform and approximately log-uniform distributions over nonzero support sizes. This mixture favors small supports without excluding larger ones. In the version with children of $Y$, selected nonparent covariates are regenerated using $Y$ and, optionally, other covariates. The selected covariates are therefore children of $Y$, and the optional inputs are their other parents. Further covariates may be generated as descendants of the children. Under the faithfulness assumption of \cref{sec:prior-labels}, the conditional dependence set contains the parents of $Y$, its children, and the other parents of its children. More distant descendants need not be included because the intervening child is observed. The $X$-first component used to train CD-PFN omits these extensions.


\subsection{Training labels and their interpretation}\label{sec:prior-labels}

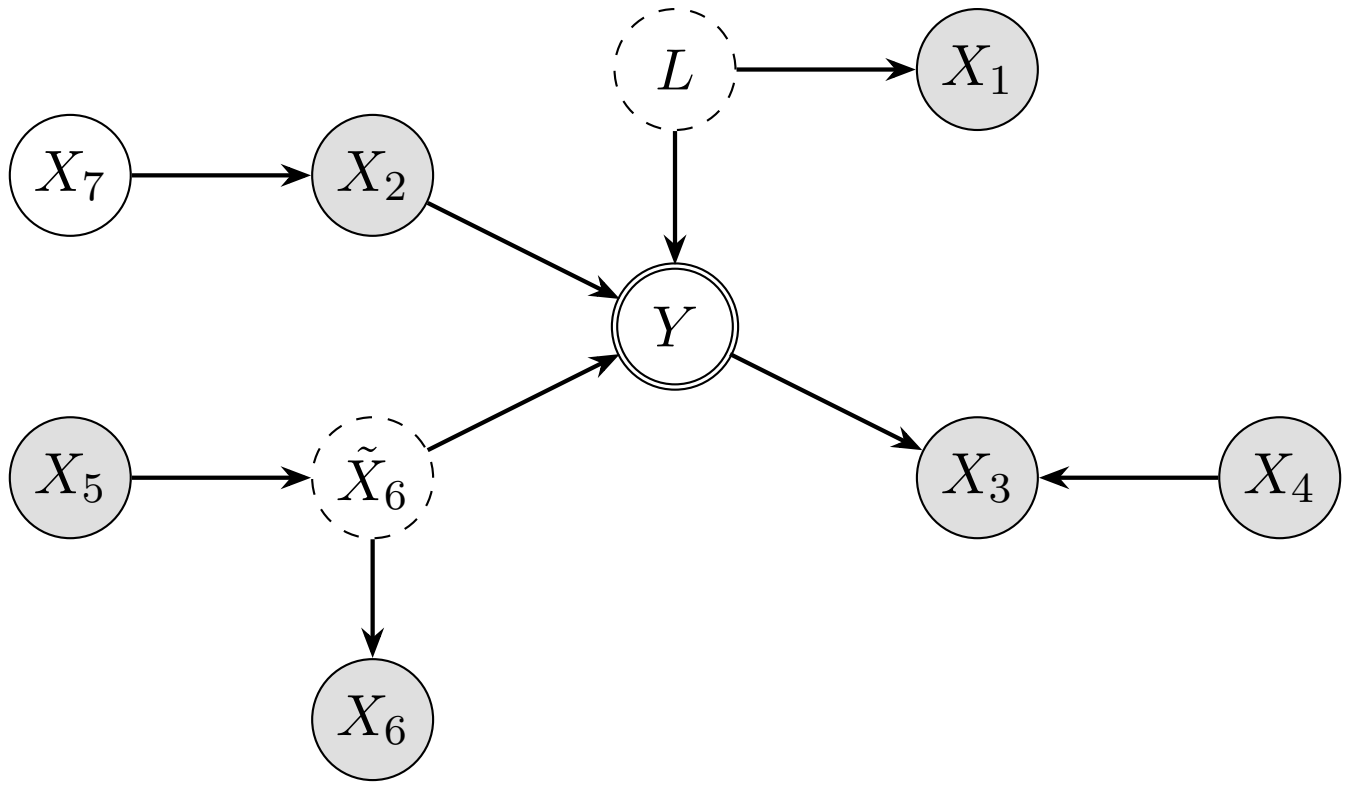
\begin{figure}[htbp]
\centering
\begin{tikzpicture}
\node[resp] (y) at (3, 0) {$Y$};
\node[lat] (l) at (3, 1.7) {$L$};
\node[obs, cd] (x1) at (5, 1.7) {$X_1$};
\node[obs, cd] (x2) at (1, 1) {$X_2$};
\node[obs] (x7) at (-1, 1) {$X_7$};
\node[lat] (z6) at (1, -1) {$\tilde X_6$};
\node[obs, cd] (x6) at (1, -2.6) {$X_6$};
\node[obs, cd] (x5) at (-1, -1) {$X_5$};
\node[obs, cd] (x3) at (5, -1) {$X_3$};
\node[obs, cd] (x4) at (7, -1) {$X_4$};
\draw[edge] (l) -- (y);
\draw[edge] (l) -- (x1);
\draw[edge] (x7) -- (x2);
\draw[edge] (x2) -- (y);
\draw[edge] (x5) -- (z6);
\draw[edge] (z6) -- (y);
\draw[edge] (z6) -- (x6);
\draw[edge] (y) -- (x3);
\draw[edge] (x4) -- (x3);
\end{tikzpicture}
\caption{\textit{Ways a covariate can be conditionally dependent on the response}. A structural causal model with covariates $X_1,\dots,X_7$ and a latent variable $L$; dashed nodes are unobserved, and $X_6$ is a coarse measurement of $\tilde X_6$. Shaded covariates are conditionally dependent on $Y$ given the others. $X_2$ is a parent, as is $X_6$ through its underlying variable $\tilde X_6$; $X_3$ is a child, and $X_4$ another parent of that child. $X_1$ shares the unobserved cause $L$ with $Y$, and $X_5$ acts through $\tilde X_6$, which is observed only coarsely; neither is a parent or a child. $X_7$ is not conditionally dependent on $Y$, because its only path to $Y$ passes through the observed $X_2$.}
\label{fig:cd-graph}
\end{figure}

For graph-based draws, the simulator computes d-separation in an augmented graph. Each coarse measurement is represented as an observed child of its underlying continuous variable, which remains latent unless observed separately. Conditioning on the measurement therefore differs from conditioning on the underlying variable. Parent and child labels use direct edges between the underlying nodes, as in the main text. \cref{fig:cd-graph} illustrates the resulting conditional dependence labels.

The conditional dependence label is one when the observed covariate and response are d-connected given the other observed covariates, meaning that some path between them has no conditioned noncollider and has every collider conditioned on or with a conditioned descendant; otherwise they are d-separated \citepS{pearl1988}. D-separation implies conditional independence under the simulated structural models; the labels additionally assume faithfulness, so that d-connection implies conditional dependence. Root nodes are pure noise, every nonroot node has independent noise with weight at least 0.05, and edges found inactive on reference observations are removed, making violations unlikely. In $X$-first draws without children, no covariate is a function of the others, and the simulator verifies after discretization that each selected covariate remains functionally active, redrawing the mechanism otherwise. Thus, the labels $\1(j\in S)$ equal $\gamma_j(P_M)$ almost surely. Parent and child labels are graph properties and require no faithfulness assumption.


\section{Conditional dependence simulations}\label{sec:cd-simulations}



\subsection{Data}\label{sec:cd-data}


\begin{table}[htbp]
\centering
\small
\begin{tabular}{@{}lccccc@{}}
\toprule
Study & $n$ & $p$ & $\lvert S\rvert$ & Parents & Children \\
\midrule
Conditional dependence & 100, \textbf{250}, 800 & 50, \textbf{200}, 500, 1000 & 5, \textbf{10}, 25 & \textemdash & \textemdash \\
Parents and children & 100, \textbf{250}, 800 & 20, \textbf{50}, 100 & \textemdash & 0, \textbf{3}, 6 & 0, \textbf{3}, 6 \\
\bottomrule
\end{tabular}
\caption{\textit{Simulation study parameters}. Each setting changes at most one baseline value (bold) to another value listed. The value $p=1000$ is run only on TCGA, and the simulated DAG for the parent and child study only at baseline.}
\label{tab:sim-settings}
\end{table}

The TCGA data consist of pan-kidney RNA-seq measurements from LinkedOmics \citepS{weinstein2013,vasaikar2018}, restricted to the 2000 most variable genes in 889 samples; all covariates are continuous. Bioresponse (OpenML 4134; \citealpS{vanschoren2014}) contains molecular descriptors for 3751 molecules, mostly counts or substructure indicators; its response is discarded. Constant, duplicated, and nearly duplicated columns, and columns that are exact functions of others, are removed without reference to a response. Each trial draws $n$ rows, then $p$ columns that are nonconstant and nonredundant within those rows, and standardizes them.


\subsection{Responses}\label{sec:cd-responses}

\begin{figure}[htbp]
\centering
\includegraphics[width=\textwidth]{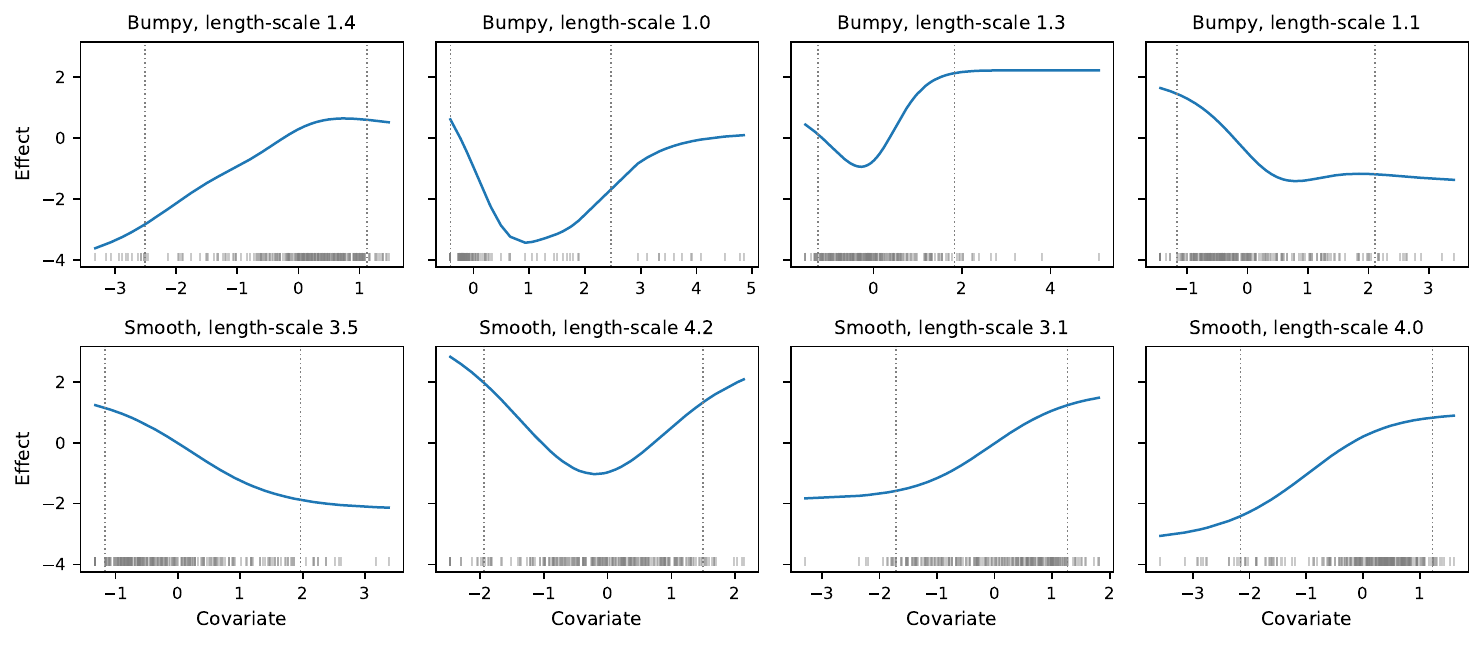}
\caption{\textit{Examples of the functions $g_j$}. Standardized functions $\tilde g_j$ of eight TCGA covariates in one sample of 250 observations, with bumpy (top) and smooth (bottom) length-scales. Dotted lines mark the 5th and 95th percentiles of the covariate, beyond which the effect flattens, and ticks at the bottom mark the observed values.}
\label{fig:cd-gp-effects}
\end{figure}

The active set $S$ is drawn uniformly among subsets of the specified size. For each $j\in S$, we draw a function $g_j$ from a zero-mean Gaussian process with a squared-exponential kernel evaluated at
\begin{align*}
c_j(x)=m+h\tanh\left(\frac{x-m}{h}\right),\qquad m=\frac{q_{95}+q_{05}}{2},\quad h=\frac{q_{95}-q_{05}}{2},
\end{align*}
which bends the standardized covariate smoothly to a constant beyond its 5th and 95th percentiles. Length-scales are drawn with equal probability from $\operatorname{Unif}(1,1.5)$ or $\operatorname{Unif}(3,5)$, producing bumpy or smooth effects (\cref{fig:cd-gp-effects}). Two disjoint pairs of active covariates contribute interactions $h_{ab}$ drawn from a two-dimensional Gaussian process. With each term standardized, as marked by a tilde, and weights $|w|\sim\operatorname{Unif}(0.5,1.5)$ with random signs,
\begin{align*}
f=\sum_{j\in S}w_j\,\tilde g_j+\sum_{(a,b)}w_{ab}\,\tilde h_{ab}.
\end{align*}
Regression responses are $y=f+\varepsilon$ with Gaussian noise at a signal-to-noise ratio of one, and classification responses are Bernoulli with probability $(1+e^{-3\tilde f})^{-1}$. Since $y$ depends on the covariates only through $X_S$, $S$ is the conditional dependence set. Two other response families are run on TCGA at baseline: linear effects without interactions and strongly nonlinear effects, with each $g_j$ a Gaussian bump, a pair of bumps, a U-shape, or an oscillation of the covariate's normal scores, its ranks mapped to standard Gaussian quantiles.

\subsection{Design and methods}\label{sec:cd-methods}


The baseline is $n=250$, $p=200$, and $\lvert S\rvert=10$. We vary $n\in\{100,800\}$, $p\in\{50,500\}$, and $\lvert S\rvert\in\{5,25\}$ one at a time for both response types and datasets, and also run $p=1000$ for both response types on TCGA. Each setting has 100 trials, and all methods receive the same datasets.

\begin{table}[htbp]
\centering
\small
\begin{tabular}{l>{\raggedright\arraybackslash}p{3.1cm}>{\raggedright\arraybackslash}p{8.1cm}}
\toprule
Method & Software & Settings \\
\midrule
Spike-and-slab & BAS (R; \citealpS{clyde2025}) & Beta-binomial(1, 1) model prior; MCMC with 20{,}000 iterations; logistic model for binary responses. Illustrative simulation: Zellner--Siow slab, Beta(1, 9) prior on the inclusion rate, 100{,}000 iterations \\
\addlinespace
SuSiE & susieR (R) & $L=30$ single effects; binary responses treated as continuous \\
\addlinespace
DART & BART (R) & Sparse prior; 1000 burn-in and 1000 posterior draws (5000 and 2000 in the preview simulation); probit model for binary responses \\
\addlinespace
BKMR & bkmr (R) & Beta(1, 9) prior on the inclusion rate; 5000 iterations, second half kept; preview simulation only \\
\addlinespace
TabPFN & tabpfn (Python) & Permutation importance with 5-fold cross-fitting and 5 permutations per covariate; preview simulation only \\
\addlinespace
Random forest & scikit-learn & 300 trees; impurity importance; jittered covariates \\
\addlinespace
XGBoost & xgboost & 300 trees of depth 4, learning rate 0.05, subsample 0.8; gain importance \\
\addlinespace
Lasso & scikit-learn & Order of entry along the path; $\ell_1$-penalized logistic path for binary responses \\
\addlinespace
Partial correlation & NumPy & Ledoit--Wolf covariance \\
\addlinespace
OLS $t$-statistics & statsmodels & Linear probability model for binary responses; only when $n>p+1$ \\
\addlinespace
Marginal correlation & NumPy & None \\
\addlinespace
Mutual information & scikit-learn & Nearest-neighbor estimator \\
\addlinespace
Model-X knockoffs & knockoff (R) & Second-order Gaussian knockoffs (ASDP); statistics from lasso coefficients (glmnet, 5-fold cross-validation) or random forest importance; knockoff+ \\
\addlinespace
IPSS & ipss (Python) & Gradient boosting selector; jittered covariates \\
\addlinespace
Benjamini--Hochberg & statsmodels & OLS $p$-values, or marginal regression $p$-values when $n\leq p+1$ \\
\bottomrule
\end{tabular}
\caption{\textit{Conditional dependence methods}. Software and settings of the methods compared with CD-PFN in the simulations and real-data studies. Settings not listed are package defaults. Jittered covariates have uniform noise added to each column with fewer than 15 distinct values, which reduces the bias of tree-based importance toward covariates with many distinct values \protect\citepS{li2026}.}
\label{tab:methods}
\end{table}

\begin{table}[htbp]
\centering
\small
\begin{tabular}{l>{\raggedright\arraybackslash}p{3.1cm}>{\raggedright\arraybackslash}p{8.1cm}}
\toprule
Method & Software & Settings \\
\midrule
AVICI & avici (Python) & Pretrained \texttt{scm-v0} model \\
\addlinespace
BiDAG & BiDAG (R; \citealpS{kuipers2022}) & BGe score \citepS{geiger2002}; iterative MCMC to find the search space, then partition MCMC; posterior edge probabilities \\
\addlinespace
DAGMA & dagma (Python; \citealpS{bello2022}) & Linear, least-squares loss; scores are absolute edge weights, and selections are edges with absolute weight above 0.3 \\
\addlinespace
PC & causal-learn (Python) & Level 0.05; Fisher $z$ or randomized conditional independence tests \citepS{strobl2019} \\
\addlinespace
GES & causal-learn (Python) & BIC score \\
\addlinespace
MMHC & bnlearn (R) & None \\
\bottomrule
\end{tabular}
\caption{\textit{Parent and child methods}. Software and settings of the methods compared with PC-PFN. Settings not listed are package defaults.}
\label{tab:methods-pc}
\end{table}

CD-PFN is applied as pretrained in one CPU forward pass. The comparison settings are given in \cref{tab:methods}. Spike-and-slab and SuSiE select covariates by the rule of \cref{sec:bvs}. The selection procedures rely on different assumptions: knockoffs control FDR for conditional dependence only when the knockoffs are valid, which second-order Gaussian knockoffs need not be for non-Gaussian covariates, while marginal $p$-values do not target conditional dependence. DART's inclusion probability is the share of posterior draws in which a covariate appears in at least one tree; because this quantity can be high for null covariates, DART is compared on ranking only. BKMR and TabPFN with permutation importance \citepS{hollmann2025}, which requires over a thousand TabPFN predictions per dataset, are too costly for the full simulations and appear only in the preview simulation (\cref{sec:illustrative-simulation}). All settings are provided in the code.


\subsection{Ranking}\label{sec:cd-results}


\cref{tab:cd-paired} gives the paired AUPRC difference between CD-PFN and the competitor with the highest mean in each setting. \cref{tab:cd-auprc-reg-tcga,tab:cd-auprc-reg-bio,tab:cd-auprc-cls-tcga,tab:cd-auprc-cls-bio} give the mean AUPRC of every method in every setting. Standard errors over the 100 trials are at most 0.015.

\begin{table}[htbp]
\centering
\begin{adjustbox}{max width=\textwidth}
\input{tables/cd_paired}
\end{adjustbox}
\caption{\textit{Paired AUPRC differences}. Mean difference between CD-PFN and the competitor with the highest mean AUPRC in each setting, with its standard error in parentheses. The difference exceeds two standard errors in every setting except Bioresponse regression at $p=500$ and classification at $n=800$.}
\label{tab:cd-paired}
\end{table}

\begin{table}[htbp]
\centering
\begin{adjustbox}{max width=\textwidth}
\input{tables/cd_auprc_regression_tcga_kidney}
\end{adjustbox}
\caption{\textit{AUPRC, regression, TCGA}. Bold marks the best method.}
\label{tab:cd-auprc-reg-tcga}
\end{table}

\begin{table}[htbp]
\centering
\begin{adjustbox}{max width=\textwidth}
\input{tables/cd_auprc_regression_bioresponse}
\end{adjustbox}
\caption{\textit{AUPRC, regression, Bioresponse}. Bold marks the best method.}
\label{tab:cd-auprc-reg-bio}
\end{table}

\begin{table}[htbp]
\centering
\begin{adjustbox}{max width=\textwidth}
\input{tables/cd_auprc_classification_tcga_kidney}
\end{adjustbox}
\caption{\textit{AUPRC, classification, TCGA}. Bold marks the best method.}
\label{tab:cd-auprc-cls-tcga}
\end{table}

\begin{table}[htbp]
\centering
\begin{adjustbox}{max width=\textwidth}
\input{tables/cd_auprc_classification_bioresponse}
\end{adjustbox}
\caption{\textit{AUPRC, classification, Bioresponse}. Bold marks the best method.}
\label{tab:cd-auprc-cls-bio}
\end{table}


\subsection{Response families}\label{sec:cd-families}


To assess sensitivity to the form of the effects, we repeat the TCGA baseline with purely linear effects without interactions and with strongly nonlinear effects such as bumps, U-shapes, and oscillations (\cref{tab:cd-families}). Under linear effects, spike-and-slab regression is correctly specified; it ranks covariates better than CD-PFN for continuous responses, and the two are tied for binary responses. Under strongly nonlinear effects, the linear methods perform poorly; CD-PFN ranks best for continuous responses and ties with random forests for binary responses. Bumps and oscillations are included among the prior's edge functions.

\begin{table}[htbp]
\centering
\begin{adjustbox}{max width=\textwidth}
\input{tables/cd_families}
\end{adjustbox}
\caption{\textit{Response families}. Mean AUPRC on TCGA at the baseline, for linear effects, the main family of the other results, and strongly nonlinear effects. Bold marks the best method in each column, including ties.}
\label{tab:cd-families}
\end{table}


\subsection{FDR control}\label{sec:cd-fdr}


\cref{tab:cd-fdr-reg-tcga,tab:cd-fdr-reg-bio,tab:cd-fdr-cls-tcga,tab:cd-fdr-cls-bio} give the empirical FDR and TPR at target level 0.1 in every setting. Across the 30 settings, CD-PFN's empirical FDR is at most 0.043 at target level 0.05 and at most 0.094 at target level 0.1. Among methods whose empirical FDR is at most the target, CD-PFN has the highest TPR in 20 settings, spike-and-slab and IPSS in 5, and SuSiE in 2, counting ties. Knockoffs with lasso and random forest statistics select few covariates in every setting.

\begin{table}[htbp]
\centering
\begin{adjustbox}{max width=\textwidth}
\input{tables/cd_fdr_regression_tcga_kidney}
\end{adjustbox}
\caption{\textit{Empirical FDR and TPR at target level 0.1, regression, TCGA}. Bold marks the highest TPR, including ties, among methods whose empirical FDR is at most 0.1. $^*$Marginal tests.}
\label{tab:cd-fdr-reg-tcga}
\end{table}

\begin{table}[htbp]
\centering
\begin{adjustbox}{max width=\textwidth}
\input{tables/cd_fdr_regression_bioresponse}
\end{adjustbox}
\caption{\textit{Empirical FDR and TPR at target level 0.1, regression, Bioresponse}. Bold marks the highest TPR, including ties, among methods whose empirical FDR is at most 0.1. $^*$Marginal tests.}
\label{tab:cd-fdr-reg-bio}
\end{table}

\begin{table}[htbp]
\centering
\begin{adjustbox}{max width=\textwidth}
\input{tables/cd_fdr_classification_tcga_kidney}
\end{adjustbox}
\caption{\textit{Empirical FDR and TPR at target level 0.1, classification, TCGA}. Bold marks the highest TPR, including ties, among methods whose empirical FDR is at most 0.1. $^*$Marginal tests.}
\label{tab:cd-fdr-cls-tcga}
\end{table}

\begin{table}[htbp]
\centering
\begin{adjustbox}{max width=\textwidth}
\input{tables/cd_fdr_classification_bioresponse}
\end{adjustbox}
\caption{\textit{Empirical FDR and TPR at target level 0.1, classification, Bioresponse}. Bold marks the highest TPR, including ties, among methods whose empirical FDR is at most 0.1. $^*$Marginal tests.}
\label{tab:cd-fdr-cls-bio}
\end{table}

\FloatBarrier


\section{Parent and child simulations}\label{sec:pc-simulations}



\subsection{Data}


The main design uses the TCGA covariates of \cref{sec:cd-data}. In each trial, $\lvert\mathrm{pa}\rvert$ parents and $\lvert\mathrm{ch}\rvert$ children are chosen among the sampled columns. The response is generated from its parents using the family of \cref{sec:cd-responses} at a signal-to-noise ratio of one, or as noise when there are no parents. Each child column $x_c$ is replaced by the standardized value of $x_c+b\,g(y)$, where $g$ is a Gaussian process draw as in \cref{sec:cd-responses} and $b$ is scaled so that $\operatorname{Var}\{b\,g(y)\}=0.2\operatorname{Var}(x_c)$. At baseline, each of the three parents explains about a sixth of the response's variance, and the response explains about a sixth of each child's variance.

The second design uses a random linear Gaussian DAG with the response in the middle of the causal order, $\lvert\mathrm{pa}\rvert$ parents, $\lvert\mathrm{ch}\rvert$ children, other edges drawn independently with probability $2/(p+1)$, weights $|w|\sim\operatorname{Unif}(0.5,1.5)$, and a signal-to-noise ratio of one at each node. All variables are standardized so that marginal variances do not reveal the causal order \citepS{reisach2021}.


\subsection{Design and methods}


The baseline is $n=250$, $p=50$, with three parents and three children. In the TCGA design, we vary $n\in\{100,800\}$, $p\in\{20,100\}$, and the numbers of parents and children in $\{0,6\}$ one at a time. The simulated DAG is run only at baseline. Each setting has 100 trials. PC-PFN, AVICI \citepS{lorch2022}, and BiDAG select from their parent and child probabilities using the rule of \cref{sec:bvs}. DAGMA selects edges into and out of the response whose absolute weight exceeds its default threshold. \cref{tab:methods-pc} gives all method settings. For PC, GES, and MMHC, directed edges at the response define its parents and children. Undirected edges count as neither, so these methods abstain when the Markov equivalence class, the set of DAGs implying the same conditional independencies \citepS{spirtes2000}, leaves a direction unresolved, whereas the pretrained models always return directional probabilities. The level of PC's tests is not an FDR target. PC with kernel tests and GES took over a minute per dataset on TCGA and were run only on the simulated DAG. BiDAG took about 7 minutes per dataset at the TCGA baseline and over 40 minutes at $n=800$ and $p=100$, where it was omitted.


\subsection{Results}


\cref{tab:pc-cells-auprc} gives the AUPRC of methods that return scores, while \cref{tab:pc-cells-parent,tab:pc-cells-child} give the empirical FDR and TPR of every method. FDR is averaged over all trials and TPR over trials with at least one true parent or child. In settings without parents or children, FDR is therefore the proportion of trials with any selection.

Simulated causal orders can leak through $R^2$-sortability, the tendency of variables later in the causal order to be more predictable from the others \citepS{reisach2023}. A rule that calls a true neighbor a parent when its $R^2$ given all other variables is below that of the response orients 41\% to 51\% of neighbors correctly across settings, near chance, whereas PC-PFN orients 85\% to 95\% correctly.

\begin{table}[htbp]
\centering
\begin{adjustbox}{max width=\textwidth}
\input{tables/pc_cells_auprc}
\end{adjustbox}
\caption{\textit{AUPRC of the methods that return scores}. Means over trials with at least one true parent or child; bold marks the highest, including ties.}
\label{tab:pc-cells-auprc}
\end{table}

\begin{table}[htbp]
\centering
\begin{adjustbox}{max width=\textwidth}
\input{tables/pc_cells_parent}
\end{adjustbox}
\caption{\textit{Parents: empirical FDR and TPR}. Target level 0.1. Settings vary one parameter from the TCGA baseline. Bold marks the highest TPR, including ties, among methods whose empirical FDR is at most 0.1.}
\label{tab:pc-cells-parent}
\end{table}

\begin{table}[htbp]
\centering
\begin{adjustbox}{max width=\textwidth}
\input{tables/pc_cells_child}
\end{adjustbox}
\caption{\textit{Children: empirical FDR and TPR}. Target level 0.1. Bold marks the highest TPR, including ties, among methods whose empirical FDR is at most 0.1.}
\label{tab:pc-cells-child}
\end{table}

\FloatBarrier


\section{Real data}\label{sec:real-data}


Both PFNs are applied as pretrained at target FDR level 0.1. CD-PFN is compared with the methods of \cref{sec:cd-methods} and PC-PFN with those of \cref{sec:pc-simulations}. For each study, the covariates and established relations were fixed before the data were examined.


\subsection{Biomarkers}\label{sec:rd-biomarkers}


Each task uses a tumor status as the response and gene expression as covariates. Expression measurements come from LinkedOmics \citepS{vasaikar2018}, and status labels, measured independently of gene expression, come from cBioPortal \citepS{cerami2012}. The established biomarkers are ESR1 and PGR for estrogen and progesterone receptor status in breast cancer, which they encode and which define the statuses clinically \citepS{allison2020}; ERBB2 for HER2 status \citepS{wolff2018}; all three for triple-negative status; CDH1 for lobular histology, whose loss is characteristic of invasive lobular carcinoma \citepS{berx1995}; CDKN2A for HPV status in head and neck cancer, whose product p16 is overexpressed in HPV-positive tumors and is the clinical surrogate for HPV in oropharyngeal cancer \citepS{lewis2018}; and MLH1 for microsatellite instability in stomach, colorectal, and endometrial cancer, where it is silenced in most MSI tumors \citepS{herman1998,tcga2013ucec,tcga2014stad}.

The covariates, chosen without the response, are the expression levels of the 456 measured genes in the MSK-IMPACT cancer panel \citepS{cheng2015}. None of these cohorts was used in training. \cref{tab:rd-biomarkers} gives the results. To assess sample-size sensitivity, we also run each single-biomarker task on 20 response-stratified subsamples of 50, 100, and 200 patients. CD-PFN selects the biomarker in 32\%, 71\%, and 86\% of subsamples, respectively, with at most 0.1 other selections on average. The corresponding rates are 17\%, 45\%, and 76\% for spike-and-slab; 65\%, 84\%, and 92\% for SuSiE, with up to 1.2 other selections; and 57\%, 87\%, and 99\% for IPSS, with up to 4.7 other selections.

\begin{table}[htbp]
\centering
\begin{adjustbox}{max width=\textwidth}
\input{tables/rd_biomarkers}
\end{adjustbox}
\caption{\textit{Biomarkers}. Established biomarker rank among the covariates, shown in bold when it is selected at target FDR level 0.1. Tied scores share the average of their ranks. DART, lasso, and correlation are used for ranking only.}
\label{tab:rd-biomarkers}
\end{table}


\subsection{Cell lines}\label{sec:rd-celllines}


Responses are either the CRISPR knockout effect of a gene in 1103 DepMap cell lines \citepS{depmap2024,tsherniak2017} or the area under the dose--response curve of a drug in about 700 GDSC lines \citepS{iorio2016}. The covariates, chosen without the response, are expression levels for the 504 genes of the 505-gene MSK-IMPACT panel measured in DepMap, together with indicators of hotspot and damaging mutations in panel genes mutated in enough lines.

The established biomarkers fall into five mechanistic groups. The first consists of paralog-loss biomarkers: SMARCA4 for SMARCA2 dependency, ARID1A for ARID1B, and STAG2 for STAG1. When one of two redundant paralogs is lost, cells depend on the other, a synthetic lethality found in RNAi and CRISPR screens \citepS{hoffman2014,helming2014,vanderlelij2017}. The second is MLH1 for WRN: cells with microsatellite instability, usually caused by MLH1 silencing, depend on the WRN helicase \citepS{chan2019}. The third uses an oncogene's own expression or mutation as a biomarker for dependence on that oncogene.

The fourth pairs each drug with its molecular target or an established sensitivity biomarker: lapatinib with ERBB2, fulvestrant with ESR1, dabrafenib with BRAF, nutlin-3a with TP53, alpelisib with PIK3CA, afatinib with EGFR, and venetoclax with BCL2. The fifth covers drugs with several established biomarkers: BRAF, KRAS, and NRAS mutations for the MEK inhibitors trametinib, selumetinib, and PD0325901, which act downstream of RAS and RAF; PIK3CA mutation and PTEN loss for the AKT inhibitors ipatasertib and MK-2206; and TP53 mutation and MDM2 expression for nutlin-3a, which blocks MDM2 and requires wild-type TP53. When both expression and mutation of a gene are biomarkers, the gene counts as found if either is selected. \cref{tab:rd-cellline-single,tab:rd-cellline-multi} give the results for responses with one or several established biomarkers.

\begin{table}[htbp]
\centering
\begin{adjustbox}{max width=\textwidth}
\input{tables/rd_cellline_single}
\end{adjustbox}
\caption{\textit{Cell lines with one biomarker}. As in \cref{tab:rd-biomarkers}; a biomarker given by both expression and mutation takes the better rank of the two.}
\label{tab:rd-cellline-single}
\end{table}

\begin{table}[htbp]
\centering
\begin{adjustbox}{max width=\textwidth}
\input{tables/rd_cellline_multi}
\end{adjustbox}
\caption{\textit{Drugs with several biomarkers}. Ranks of the established biomarkers in increasing order, as in \cref{tab:rd-biomarkers}.}
\label{tab:rd-cellline-multi}
\end{table}


\subsection{Metabolic syndrome}\label{sec:rd-nhanes}


We use the 2076 adults in the fasting subsample of NHANES 2017--2018 with complete covariates. Half have metabolic syndrome under the harmonized definition \citepS{alberti2009}: at least three of a large waist, high triglycerides, low HDL cholesterol, high blood pressure, and high fasting glucose, with medication counting toward the last four. Sex determines the waist and HDL thresholds, making it the seventh defining variable. The other 58 covariates are age, ethnicity, education, and exam and laboratory measurements with at most 10\% missing; near-duplicates of the defining variables are removed. Medications enter the definition but are not covariates, so other selections need not be false. We draw 20 random samples at each $n\in\{100,200,500,1000\}$. SuSiE never selects HDL cholesterol at $n\geq500$, whereas CD-PFN selects it in 80\% of samples at $n=500$ and all samples at $n=1000$. \cref{tab:rd-nhanes} gives the full results.

\begin{table}[htbp]
\centering
\begin{adjustbox}{max width=\textwidth}
\input{tables/rd_nhanes}
\end{adjustbox}
\caption{\textit{Metabolic syndrome}. Mean number of the seven defining variables selected, with the mean number of other selections in parentheses.}
\label{tab:rd-nhanes}
\end{table}


\subsection{Exogenous variables}\label{sec:rd-exogenous}


Age, sex, and ethnicity cannot be caused by any measurement in NHANES. We use them to check direction in two ways on all 2076 participants described in \cref{sec:rd-nhanes}, about twice the largest sample size in the training prior. First, each of the 61 measurements serves in turn as the response, with the other 64 variables as covariates; age, sex, and ethnicity can then be parents of the response but not children. Second, each demographic variable serves as the response; the measurements can then be children but not parents. All variables are standardized, and selections use target FDR level 0.1 with the methods and settings of \cref{sec:pc-simulations}.

PC-PFN took under 15 seconds per dataset. PC took over 30 minutes, and GES did not finish within four hours, so neither is included. BiDAG took between one and two and a half hours per dataset and was run only with the demographic responses. \cref{tab:rd-exogenous} gives the results, and \cref{fig:rd-exogenous} shows each method's selections with age and sex as the response.

\begin{table}[htbp]
\centering
\begin{adjustbox}{max width=\textwidth}
\input{tables/rd_exogenous}
\end{adjustbox}
\caption{\textit{Direction for exogenous variables, all selections}. Demographic covariates: with each of the 61 NHANES measurements as the response, the numbers of selections of age, sex, and ethnicity as parents (correct) and as children (incorrect). Demographic responses: with each of age, sex, and ethnicity as the response, the numbers of measurements selected as children (correct) and as parents (incorrect). BiDAG is run with demographic responses only.}
\label{tab:rd-exogenous}
\end{table}


\subsection{Light tunnel}\label{sec:rd-chamber}


The light tunnel of \citeS{gamella2025} is a physical apparatus in which a light source shines through two linear polarizers onto infrared and visible light sensors at three positions, with additional LEDs beside the sensors. We use the observational reference experiment from the \texttt{causalchamber} package, in which the inputs are set at random (10{,}000 observations). The inputs are the red, green, and blue brightness of the light source, the two polarizer angles, and the brightness of six LEDs. The measurements are the current drawn by the light source and the six light-sensor readings. Variables held constant in the experiment, including sensor settings, are removed. The two angle sensors read their polarizers almost exactly and are merged with their corresponding polarizer variables. The remaining 18 variables and 37 edges in the ground-truth graph distributed with the package define the truth: inputs have only children, and measurements have only parents. Each variable serves once as the response, with the other 17 as covariates, using all 10{,}000 observations. All variables are standardized. Parent AUPRC is averaged over responses with parents, and child AUPRC over responses with children. Empirical FDR is averaged over all responses, while TPR is averaged over responses with parents or children.

\bibliographystyleS{plainnat}
\bibliographyS{refs}

\end{document}

%% file: tables/pc_baseline.tex
\begin{tabular}{lcccccccc}
\toprule
 & \multicolumn{4}{c}{TCGA (baseline)} & \multicolumn{4}{c}{Simulated DAG} \\
\cmidrule(lr){2-5}\cmidrule(lr){6-9}
 & \multicolumn{2}{c}{Parents} & \multicolumn{2}{c}{Children} & \multicolumn{2}{c}{Parents} & \multicolumn{2}{c}{Children} \\
\cmidrule(lr){2-3}\cmidrule(lr){4-5}\cmidrule(lr){6-7}\cmidrule(lr){8-9}
Method & AUPRC & FDR / TPR & AUPRC & FDR / TPR & AUPRC & FDR / TPR & AUPRC & FDR / TPR \\
\midrule
PC-PFN & $\mathbf{0.87}$ & $\mathbf{0.04}$ / $\mathbf{0.59}$ & $\mathbf{0.98}$ & $\mathbf{0.08}$ / $\mathbf{0.94}$ & $0.91$ & $0.03$ / $0.46$ & $0.94$ & $0.04$ / $0.74$ \\
BiDAG & $0.59$ & $0.28$ / $0.51$ & $0.67$ & $0.26$ / $0.61$ & $\mathbf{0.93}$ & $\mathbf{0.05}$ / $\mathbf{0.83}$ & $\mathbf{0.96}$ & $\mathbf{0.09}$ / $\mathbf{0.95}$ \\
DAGMA & $0.37$ & $0.66$ / $0.24$ & $0.44$ & $0.28$ / $0.23$ & $0.37$ & $0.48$ / $0.21$ & $0.60$ & $0.46$ / $0.58$ \\
AVICI & $0.24$ & $0.00$ / $0.00$ & $0.55$ & $0.00$ / $0.01$ & $0.50$ & $0.01$ / $0.02$ & $0.68$ & $0.00$ / $0.02$ \\
PC (Fisher $z$) & \textemdash & $0.45$ / $0.49$ & \textemdash & $0.17$ / $0.09$ & \textemdash & $0.27$ / $0.76$ & \textemdash & $0.04$ / $0.45$ \\
PC (RCIT) & \textemdash & \textemdash & \textemdash & \textemdash & \textemdash & $0.23$ / $0.36$ & \textemdash & $0.10$ / $0.25$ \\
GES & \textemdash & \textemdash & \textemdash & \textemdash & \textemdash & $0.20$ / $0.82$ & \textemdash & $0.21$ / $0.94$ \\
MMHC & \textemdash & $0.25$ / $0.30$ & \textemdash & $0.34$ / $0.48$ & \textemdash & $0.13$ / $0.54$ & \textemdash & $0.21$ / $0.79$ \\
\bottomrule
\end{tabular}

%% file: tables/rd_summary.tex
\begin{tabular}{lccc}
\toprule
Method & One biomarker & Several biomarkers & Metabolic syndrome \\
\midrule
CD-PFN & $21/25$ ($1.6$) & $15/18$ ($1.6$) & $5.8/7$ ($3.0$) \\
Spike-and-slab & $22/25$ ($3.0$) & $11/18$ ($3.4$) & $4.5/7$ ($3.2$) \\
SuSiE & $23/25$ ($2.8$) & $12/18$ ($2.1$) & $4.3/7$ ($2.5$) \\
IPSS & $23/25$ ($5.9$) & $13/18$ ($6.7$) & $6.0/7$ ($6.0$) \\
BH & $16/25$ ($24.8$) & $2/18$ ($0.3$) & $3.4/7$ ($2.0$) \\
Knockoffs & $0/25$ ($0.0$) & $0/18$ ($0.0$) & $1.4/7$ ($1.3$) \\
Knockoffs (RF) & $0/25$ ($0.0$) & $0/18$ ($0.0$) & $2.1/7$ ($2.3$) \\
\bottomrule
\end{tabular}

%% file: tables/rd_chambers_pc.tex
\begin{tabular}{lcccc}
\toprule
 & \multicolumn{2}{c}{Parents} & \multicolumn{2}{c}{Children} \\
\cmidrule(lr){2-3}\cmidrule(lr){4-5}
Method & AUPRC & FDR / TPR & AUPRC & FDR / TPR \\
\midrule
PC-PFN & $0.68$ & $0.06$ / $0.44$ & $\mathbf{0.65}$ & $0.28$ / $0.12$ \\
BiDAG & $\mathbf{0.70}$ & $0.21$ / $0.51$ & $0.52$ & $0.19$ / $0.31$ \\
DAGMA & $0.37$ & $0.42$ / $0.08$ & $0.38$ & $0.32$ / $0.04$ \\
AVICI & $0.23$ & $0.00$ / $0.00$ & $0.19$ & $\mathbf{0.00}$ / $\mathbf{0.00}$ \\
PC (Fisher $z$) & \textemdash & $0.33$ / $0.23$ & \textemdash & $0.22$ / $0.04$ \\
PC (RCIT) & \textemdash & $0.31$ / $0.35$ & \textemdash & $0.11$ / $0.06$ \\
GES & \textemdash & $\mathbf{0.09}$ / $\mathbf{0.69}$ & \textemdash & $0.33$ / $0.45$ \\
MMHC & \textemdash & $0.25$ / $0.06$ & \textemdash & $0.28$ / $0.12$ \\
\bottomrule
\end{tabular}

%% file: tables/cd_paired.tex
\begin{tabular}{lcccc}
\toprule
Setting & TCGA (regression) & Bioresponse (regression) & TCGA (classification) & Bioresponse (classification) \\
\midrule
Baseline & $0.096$ ($0.012$) & $0.024$ ($0.008$) & $0.126$ ($0.011$) & $0.118$ ($0.010$) \\
$n=100$ & $0.057$ ($0.009$) & $0.018$ ($0.009$) & $0.051$ ($0.008$) & $0.066$ ($0.009$) \\
$n=800$ & $0.114$ ($0.012$) & $0.042$ ($0.008$) & $0.110$ ($0.011$) & $0.010$ ($0.007$) \\
$p=50$ & $0.073$ ($0.010$) & $0.021$ ($0.006$) & $0.101$ ($0.011$) & $0.023$ ($0.006$) \\
$p=500$ & $0.069$ ($0.012$) & $0.004$ ($0.011$) & $0.089$ ($0.012$) & $0.099$ ($0.011$) \\
$p=1000$ & $0.062$ ($0.012$) & \textemdash & $0.078$ ($0.011$) & \textemdash \\
$\lvert S\rvert=5$ & $0.124$ ($0.017$) & $0.041$ ($0.011$) & $0.119$ ($0.016$) & $0.066$ ($0.015$) \\
$\lvert S\rvert=25$ & $0.078$ ($0.006$) & $0.035$ ($0.006$) & $0.084$ ($0.006$) & $0.069$ ($0.005$) \\
\bottomrule
\end{tabular}

%% file: tables/cd_auprc_regression_tcga_kidney.tex
\begin{tabular}{lcccccccc}
\toprule
Method & Baseline & $n=100$ & $n=800$ & $p=50$ & $p=500$ & $p=1000$ & $\lvert S\rvert=5$ & $\lvert S\rvert=25$ \\
\midrule
CD-PFN & $\mathbf{0.57}$ & $\mathbf{0.32}$ & $\mathbf{0.84}$ & $\mathbf{0.78}$ & $\mathbf{0.42}$ & $\mathbf{0.35}$ & $\mathbf{0.72}$ & $\mathbf{0.43}$ \\
Spike-and-slab & $0.47$ & $0.27$ & $0.72$ & $0.71$ & $0.35$ & $0.29$ & $0.60$ & $0.36$ \\
SuSiE & $0.39$ & $0.23$ & $0.66$ & $0.64$ & $0.26$ & $0.21$ & $0.55$ & $0.31$ \\
DART & $0.39$ & $0.14$ & $0.69$ & $0.62$ & $0.24$ & $0.08$ & $0.49$ & $0.31$ \\
Random forest & $0.39$ & $0.22$ & $0.64$ & $0.63$ & $0.26$ & $0.20$ & $0.56$ & $0.31$ \\
Lasso & $0.33$ & $0.23$ & $0.51$ & $0.58$ & $0.23$ & $0.16$ & $0.43$ & $0.31$ \\
Partial correlation & $0.33$ & $0.22$ & $0.65$ & $0.67$ & $0.18$ & $0.15$ & $0.43$ & $0.31$ \\
OLS & $0.16$ & \textemdash & $0.64$ & $0.64$ & \textemdash & \textemdash & $0.18$ & $0.19$ \\
XGBoost & $0.29$ & $0.17$ & $0.57$ & $0.62$ & $0.14$ & $0.07$ & $0.45$ & $0.26$ \\
Marginal correlation & $0.22$ & $0.17$ & $0.28$ & $0.45$ & $0.15$ & $0.10$ & $0.31$ & $0.23$ \\
Mutual information & $0.13$ & $0.11$ & $0.20$ & $0.37$ & $0.06$ & $0.03$ & $0.20$ & $0.17$ \\
\bottomrule
\end{tabular}

%% file: tables/cd_auprc_regression_bioresponse.tex
\begin{tabular}{lccccccc}
\toprule
Method & Baseline & $n=100$ & $n=800$ & $p=50$ & $p=500$ & $\lvert S\rvert=5$ & $\lvert S\rvert=25$ \\
\midrule
CD-PFN & $\mathbf{0.67}$ & $\mathbf{0.36}$ & $\mathbf{0.90}$ & $\mathbf{0.85}$ & $\mathbf{0.55}$ & $\mathbf{0.80}$ & $\mathbf{0.50}$ \\
Spike-and-slab & $0.65$ & $0.34$ & $0.86$ & $0.83$ & $\mathbf{0.55}$ & $0.76$ & $0.47$ \\
SuSiE & $0.60$ & $0.32$ & $0.85$ & $0.80$ & $0.50$ & $0.74$ & $0.43$ \\
DART & $0.39$ & $0.14$ & $0.77$ & $0.64$ & $0.27$ & $0.47$ & $0.31$ \\
Random forest & $0.46$ & $0.25$ & $0.70$ & $0.69$ & $0.34$ & $0.66$ & $0.32$ \\
Lasso & $0.53$ & $0.30$ & $0.73$ & $0.76$ & $0.41$ & $0.66$ & $0.41$ \\
Partial correlation & $0.54$ & $0.29$ & $0.80$ & $0.81$ & $0.33$ & $0.62$ & $0.46$ \\
OLS & $0.13$ & \textemdash & $0.69$ & $0.78$ & \textemdash & $0.17$ & $0.19$ \\
XGBoost & $0.41$ & $0.21$ & $0.61$ & $0.70$ & $0.24$ & $0.56$ & $0.29$ \\
Marginal correlation & $0.37$ & $0.27$ & $0.46$ & $0.62$ & $0.28$ & $0.49$ & $0.32$ \\
Mutual information & $0.20$ & $0.15$ & $0.30$ & $0.45$ & $0.11$ & $0.32$ & $0.19$ \\
\bottomrule
\end{tabular}

%% file: tables/cd_auprc_classification_tcga_kidney.tex
\begin{tabular}{lcccccccc}
\toprule
Method & Baseline & $n=100$ & $n=800$ & $p=50$ & $p=500$ & $p=1000$ & $\lvert S\rvert=5$ & $\lvert S\rvert=25$ \\
\midrule
CD-PFN & $\mathbf{0.58}$ & $\mathbf{0.29}$ & $\mathbf{0.81}$ & $\mathbf{0.77}$ & $\mathbf{0.45}$ & $\mathbf{0.34}$ & $\mathbf{0.69}$ & $\mathbf{0.42}$ \\
Spike-and-slab & $0.45$ & $0.24$ & $0.70$ & $0.67$ & $0.36$ & $0.26$ & $0.57$ & $0.34$ \\
SuSiE & $0.37$ & $0.21$ & $0.61$ & $0.60$ & $0.27$ & $0.19$ & $0.49$ & $0.31$ \\
DART & $0.38$ & $0.16$ & $0.69$ & $0.67$ & $0.15$ & $0.09$ & $0.55$ & $0.31$ \\
Random forest & $0.36$ & $0.22$ & $0.49$ & $0.60$ & $0.23$ & $0.18$ & $0.49$ & $0.28$ \\
Lasso & $0.33$ & $0.20$ & $0.41$ & $0.54$ & $0.21$ & $0.15$ & $0.39$ & $0.28$ \\
Partial correlation & $0.34$ & $0.19$ & $0.61$ & $0.63$ & $0.21$ & $0.13$ & $0.40$ & $0.30$ \\
OLS & $0.16$ & \textemdash & $0.58$ & $0.61$ & \textemdash & \textemdash & $0.16$ & $0.19$ \\
XGBoost & $0.17$ & $0.10$ & $0.38$ & $0.53$ & $0.06$ & $0.04$ & $0.22$ & $0.19$ \\
Marginal correlation & $0.23$ & $0.18$ & $0.23$ & $0.44$ & $0.14$ & $0.10$ & $0.30$ & $0.21$ \\
Mutual information & $0.16$ & $0.13$ & $0.21$ & $0.40$ & $0.08$ & $0.05$ & $0.25$ & $0.19$ \\
\bottomrule
\end{tabular}

%% file: tables/cd_auprc_classification_bioresponse.tex
\begin{tabular}{lccccccc}
\toprule
Method & Baseline & $n=100$ & $n=800$ & $p=50$ & $p=500$ & $\lvert S\rvert=5$ & $\lvert S\rvert=25$ \\
\midrule
CD-PFN & $\mathbf{0.64}$ & $\mathbf{0.38}$ & $\mathbf{0.88}$ & $\mathbf{0.83}$ & $\mathbf{0.51}$ & $\mathbf{0.74}$ & $\mathbf{0.47}$ \\
Spike-and-slab & $0.50$ & $0.31$ & $0.87$ & $0.81$ & $0.40$ & $0.67$ & $0.35$ \\
SuSiE & $0.52$ & $0.31$ & $0.77$ & $0.75$ & $0.41$ & $0.60$ & $0.39$ \\
DART & $0.43$ & $0.21$ & $0.75$ & $0.68$ & $0.22$ & $0.53$ & $0.32$ \\
Random forest & $0.34$ & $0.23$ & $0.50$ & $0.60$ & $0.24$ & $0.46$ & $0.29$ \\
Lasso & $0.46$ & $0.28$ & $0.67$ & $0.72$ & $0.34$ & $0.57$ & $0.36$ \\
Partial correlation & $0.46$ & $0.30$ & $0.71$ & $0.77$ & $0.27$ & $0.50$ & $0.40$ \\
OLS & $0.14$ & \textemdash & $0.62$ & $0.72$ & \textemdash & $0.15$ & $0.18$ \\
XGBoost & $0.26$ & $0.17$ & $0.53$ & $0.62$ & $0.15$ & $0.41$ & $0.24$ \\
Marginal correlation & $0.36$ & $0.26$ & $0.43$ & $0.60$ & $0.25$ & $0.45$ & $0.29$ \\
Mutual information & $0.16$ & $0.11$ & $0.25$ & $0.41$ & $0.10$ & $0.26$ & $0.19$ \\
\bottomrule
\end{tabular}

%% file: tables/cd_families.tex
\begin{tabular}{lcccccc}
\toprule
 & \multicolumn{3}{c}{Regression} & \multicolumn{3}{c}{Classification} \\
\cmidrule(lr){2-4}\cmidrule(lr){5-7}
Method & Linear & Main & Strongly nonlinear & Linear & Main & Strongly nonlinear \\
\midrule
CD-PFN & $0.67$ & $\mathbf{0.57}$ & $\mathbf{0.40}$ & $\mathbf{0.66}$ & $\mathbf{0.58}$ & $\mathbf{0.38}$ \\
Spike-and-slab & $\mathbf{0.73}$ & $0.47$ & $0.17$ & $\mathbf{0.66}$ & $0.45$ & $0.18$ \\
SuSiE & $0.56$ & $0.39$ & $0.16$ & $0.51$ & $0.37$ & $0.16$ \\
DART & $0.39$ & $0.39$ & $0.31$ & $0.42$ & $0.38$ & $0.31$ \\
Random forest & $0.37$ & $0.39$ & $0.37$ & $0.33$ & $0.36$ & $\mathbf{0.38}$ \\
Lasso & $0.46$ & $0.33$ & $0.15$ & $0.43$ & $0.33$ & $0.15$ \\
\bottomrule
\end{tabular}

%% file: tables/cd_fdr_regression_tcga_kidney.tex
\begin{tabular}{lcccccccc}
\toprule
Method & Baseline & $n=100$ & $n=800$ & $p=50$ & $p=500$ & $p=1000$ & $\lvert S\rvert=5$ & $\lvert S\rvert=25$ \\
\midrule
CD-PFN & $0.02$ / $0.21$ & $0.01$ / $0.02$ & $\mathbf{0.07}$ / $\mathbf{0.65}$ & $\mathbf{0.05}$ / $\mathbf{0.43}$ & $\mathbf{0.03}$ / $\mathbf{0.13}$ & $0.04$ / $0.09$ & $0.02$ / $0.46$ & $0.03$ / $0.04$ \\
Spike-and-slab & $0.08$ / $0.18$ & $0.06$ / $0.03$ & $0.13$ / $0.57$ & $0.06$ / $0.35$ & $0.05$ / $0.12$ & $0.04$ / $0.10$ & $0.07$ / $0.38$ & $0.07$ / $0.05$ \\
SuSiE & $0.16$ / $0.17$ & $0.09$ / $0.03$ & $0.12$ / $0.51$ & $0.07$ / $0.27$ & $0.28$ / $0.12$ & $0.27$ / $0.10$ & $0.12$ / $0.38$ & $0.21$ / $0.04$ \\
IPSS & $\mathbf{0.10}$ / $\mathbf{0.27}$ & $\mathbf{0.05}$ / $\mathbf{0.04}$ & $0.20$ / $0.71$ & $0.13$ / $0.39$ & $0.12$ / $0.17$ & $\mathbf{0.10}$ / $\mathbf{0.12}$ & $\mathbf{0.09}$ / $\mathbf{0.50}$ & $\mathbf{0.08}$ / $\mathbf{0.08}$ \\
Knockoffs & $0.00$ / $0.01$ & $0.00$ / $0.00$ & $0.03$ / $0.04$ & $0.01$ / $0.02$ & $0.00$ / $0.00$ & \textemdash & $0.01$ / $0.01$ & $0.00$ / $0.00$ \\
Knockoffs (random forest) & $0.00$ / $0.00$ & $0.00$ / $0.00$ & $0.03$ / $0.06$ & $0.00$ / $0.00$ & $0.00$ / $0.00$ & \textemdash & $0.00$ / $0.00$ & $0.00$ / $0.00$ \\
Benjamini--Hochberg & $0.12$ / $0.02$ & $0.83^*$ / $0.48$ & $0.13$ / $0.45$ & $0.10$ / $0.33$ & $0.96^*$ / $0.67$ & $0.96^*$ / $0.68$ & $0.08$ / $0.03$ & $0.09$ / $0.01$ \\
\bottomrule
\end{tabular}

%% file: tables/cd_fdr_regression_bioresponse.tex
\begin{tabular}{lccccccc}
\toprule
Method & Baseline & $n=100$ & $n=800$ & $p=50$ & $p=500$ & $\lvert S\rvert=5$ & $\lvert S\rvert=25$ \\
\midrule
CD-PFN & $\mathbf{0.08}$ / $\mathbf{0.39}$ & $0.04$ / $0.05$ & $\mathbf{0.08}$ / $\mathbf{0.81}$ & $\mathbf{0.06}$ / $\mathbf{0.63}$ & $0.07$ / $0.29$ & $\mathbf{0.02}$ / $\mathbf{0.61}$ & $\mathbf{0.09}$ / $\mathbf{0.12}$ \\
Spike-and-slab & $\mathbf{0.09}$ / $\mathbf{0.39}$ & $\mathbf{0.05}$ / $\mathbf{0.06}$ & $0.13$ / $0.79$ & $0.07$ / $0.58$ & $\mathbf{0.08}$ / $\mathbf{0.30}$ & $0.06$ / $0.60$ & $0.11$ / $0.10$ \\
SuSiE & $0.09$ / $0.30$ & $0.12$ / $0.06$ & $0.09$ / $0.74$ & $0.03$ / $0.42$ & $0.12$ / $0.26$ & $0.05$ / $0.57$ & $0.12$ / $0.05$ \\
IPSS & $0.11$ / $0.24$ & $0.08$ / $0.03$ & $0.21$ / $0.74$ & $0.10$ / $0.38$ & $0.09$ / $0.15$ & $0.11$ / $0.51$ & $0.07$ / $0.04$ \\
Knockoffs & $0.00$ / $0.00$ & $0.00$ / $0.01$ & $0.02$ / $0.07$ & $0.02$ / $0.06$ & $0.02$ / $0.02$ & $0.00$ / $0.00$ & $0.01$ / $0.01$ \\
Knockoffs (random forest) & $0.00$ / $0.00$ & $0.00$ / $0.00$ & $0.01$ / $0.01$ & $0.00$ / $0.00$ & $0.01$ / $0.00$ & $0.00$ / $0.00$ & $0.00$ / $0.00$ \\
Benjamini--Hochberg & $0.09$ / $0.01$ & $0.70^*$ / $0.37$ & $0.18$ / $0.54$ & $0.10$ / $0.51$ & $0.92^*$ / $0.64$ & $0.16$ / $0.03$ & $0.10$ / $0.01$ \\
\bottomrule
\end{tabular}

%% file: tables/cd_fdr_classification_tcga_kidney.tex
\begin{tabular}{lcccccccc}
\toprule
Method & Baseline & $n=100$ & $n=800$ & $p=50$ & $p=500$ & $p=1000$ & $\lvert S\rvert=5$ & $\lvert S\rvert=25$ \\
\midrule
CD-PFN & $\mathbf{0.00}$ / $\mathbf{0.19}$ & $0.00$ / $0.01$ & $\mathbf{0.05}$ / $\mathbf{0.61}$ & $\mathbf{0.03}$ / $\mathbf{0.31}$ & $\mathbf{0.01}$ / $\mathbf{0.13}$ & $\mathbf{0.01}$ / $\mathbf{0.08}$ & $\mathbf{0.02}$ / $\mathbf{0.41}$ & $0.02$ / $0.03$ \\
Spike-and-slab & $\mathbf{0.07}$ / $\mathbf{0.19}$ & $0.02$ / $0.02$ & $0.14$ / $0.56$ & $0.11$ / $0.35$ & $0.08$ / $0.12$ & $0.04$ / $0.07$ & $0.05$ / $0.37$ & $\mathbf{0.10}$ / $\mathbf{0.04}$ \\
SuSiE & $0.15$ / $0.15$ & $\mathbf{0.07}$ / $\mathbf{0.04}$ & $0.18$ / $0.45$ & $0.09$ / $0.22$ & $0.24$ / $0.12$ & $0.27$ / $0.09$ & $0.12$ / $0.34$ & $0.22$ / $0.04$ \\
IPSS & $0.35$ / $0.38$ & $0.46$ / $0.14$ & $0.32$ / $0.73$ & $0.13$ / $0.32$ & $0.41$ / $0.28$ & $0.45$ / $0.21$ & $0.41$ / $0.61$ & $0.33$ / $0.16$ \\
Knockoffs & $0.00$ / $0.00$ & $0.00$ / $0.00$ & $0.02$ / $0.06$ & $0.02$ / $0.04$ & $0.00$ / $0.00$ & \textemdash & $0.00$ / $0.00$ & $0.00$ / $0.00$ \\
Knockoffs (random forest) & $0.00$ / $0.00$ & $0.00$ / $0.00$ & $0.01$ / $0.02$ & $0.00$ / $0.00$ & $0.00$ / $0.00$ & \textemdash & $0.00$ / $0.00$ & $0.00$ / $0.00$ \\
Benjamini--Hochberg & $0.12$ / $0.02$ & $0.75^*$ / $0.38$ & $0.14$ / $0.39$ & $0.13$ / $0.27$ & $0.96^*$ / $0.64$ & $0.98^*$ / $0.67$ & $0.10$ / $0.03$ & $0.08$ / $0.00$ \\
\bottomrule
\end{tabular}

%% file: tables/cd_fdr_classification_bioresponse.tex
\begin{tabular}{lccccccc}
\toprule
Method & Baseline & $n=100$ & $n=800$ & $p=50$ & $p=500$ & $\lvert S\rvert=5$ & $\lvert S\rvert=25$ \\
\midrule
CD-PFN & $\mathbf{0.07}$ / $\mathbf{0.31}$ & $0.01$ / $0.03$ & $\mathbf{0.09}$ / $\mathbf{0.77}$ & $\mathbf{0.06}$ / $\mathbf{0.57}$ & $\mathbf{0.04}$ / $\mathbf{0.20}$ & $\mathbf{0.04}$ / $\mathbf{0.50}$ & $\mathbf{0.08}$ / $\mathbf{0.09}$ \\
Spike-and-slab & $0.17$ / $0.28$ & $0.03$ / $0.02$ & $0.12$ / $0.76$ & $0.17$ / $0.62$ & $0.06$ / $0.17$ & $0.10$ / $0.49$ & $0.19$ / $0.07$ \\
SuSiE & $0.10$ / $0.18$ & $\mathbf{0.06}$ / $\mathbf{0.04}$ & $0.06$ / $0.54$ & $0.05$ / $0.28$ & $0.09$ / $0.17$ & $0.05$ / $0.39$ & $0.14$ / $0.03$ \\
IPSS & $0.36$ / $0.26$ & $0.53$ / $0.11$ & $0.33$ / $0.60$ & $0.08$ / $0.27$ & $0.41$ / $0.21$ & $0.40$ / $0.47$ & $0.40$ / $0.10$ \\
Knockoffs & $0.00$ / $0.00$ & $0.01$ / $0.00$ & $0.01$ / $0.03$ & $0.00$ / $0.02$ & $0.01$ / $0.01$ & $0.01$ / $0.01$ & $0.00$ / $0.00$ \\
Knockoffs (random forest) & $0.00$ / $0.00$ & $0.01$ / $0.00$ & $0.00$ / $0.00$ & $0.00$ / $0.00$ & $0.01$ / $0.00$ & $0.00$ / $0.00$ & $0.00$ / $0.00$ \\
Benjamini--Hochberg & $0.08$ / $0.01$ & $0.51^*$ / $0.29$ & $0.11$ / $0.40$ & $0.09$ / $0.38$ & $0.86^*$ / $0.54$ & $0.08$ / $0.03$ & $0.10$ / $0.00$ \\
\bottomrule
\end{tabular}

%% file: tables/pc_cells_auprc.tex
\begin{tabular}{lcccccccc}
\toprule
 & \multicolumn{4}{c}{Parents} & \multicolumn{4}{c}{Children} \\
\cmidrule(lr){2-5}\cmidrule(lr){6-9}
Setting & PC-PFN & BiDAG & DAGMA & AVICI & PC-PFN & BiDAG & DAGMA & AVICI \\
\midrule
Baseline & $\mathbf{0.87}$ & $0.59$ & $0.37$ & $0.24$ & $\mathbf{0.98}$ & $0.67$ & $0.44$ & $0.55$ \\
$n=100$ & $\mathbf{0.67}$ & $0.44$ & $0.28$ & $0.19$ & $\mathbf{0.95}$ & $0.60$ & $0.38$ & $0.36$ \\
$n=800$ & $\mathbf{0.95}$ & \textemdash & $0.35$ & $0.25$ & $\mathbf{1.00}$ & \textemdash & $0.45$ & $0.64$ \\
$p=20$ & $\mathbf{0.86}$ & $0.52$ & $0.39$ & $0.39$ & $\mathbf{0.95}$ & $0.58$ & $0.46$ & $0.71$ \\
$p=100$ & $\mathbf{0.84}$ & \textemdash & $0.30$ & $0.17$ & $\mathbf{0.99}$ & \textemdash & $0.40$ & $0.34$ \\
$\lvert\mathrm{pa}\rvert=0$ & \textemdash & \textemdash & \textemdash & \textemdash & $\mathbf{1.00}$ & $0.90$ & $0.32$ & $0.79$ \\
$\lvert\mathrm{pa}\rvert=6$ & $\mathbf{0.70}$ & $0.41$ & $0.30$ & $0.24$ & $\mathbf{0.97}$ & $0.61$ & $0.44$ & $0.63$ \\
$\lvert\mathrm{ch}\rvert=0$ & $\mathbf{0.94}$ & $0.76$ & $0.78$ & $0.35$ & \textemdash & \textemdash & \textemdash & \textemdash \\
$\lvert\mathrm{ch}\rvert=6$ & $\mathbf{0.80}$ & $0.34$ & $0.19$ & $0.19$ & $\mathbf{0.98}$ & $0.63$ & $0.51$ & $0.72$ \\
Simulated DAG & $0.91$ & $\mathbf{0.93}$ & $0.37$ & $0.50$ & $0.94$ & $\mathbf{0.96}$ & $0.60$ & $0.68$ \\
\bottomrule
\end{tabular}

%% file: tables/pc_cells_parent.tex
\begin{tabular}{lcccccccc}
\toprule
Setting & PC-PFN & BiDAG & DAGMA & AVICI & PC (Fisher $z$) & PC (RCIT) & GES & MMHC \\
\midrule
Baseline & $\mathbf{0.04}$ / $\mathbf{0.59}$ & $0.28$ / $0.51$ & $0.66$ / $0.24$ & $0.00$ / $0.00$ & $0.45$ / $0.49$ & \textemdash & \textemdash & $0.25$ / $0.30$ \\
$n=100$ & $\mathbf{0.03}$ / $\mathbf{0.14}$ & $0.29$ / $0.23$ & $0.73$ / $0.18$ & $0.01$ / $0.00$ & $0.44$ / $0.26$ & \textemdash & \textemdash & $0.34$ / $0.19$ \\
$n=800$ & $\mathbf{0.01}$ / $\mathbf{0.82}$ & \textemdash & $0.71$ / $0.16$ & $0.00$ / $0.00$ & $0.57$ / $0.55$ & \textemdash & \textemdash & $0.29$ / $0.33$ \\
$p=20$ & $\mathbf{0.03}$ / $\mathbf{0.54}$ & $0.30$ / $0.29$ & $0.63$ / $0.22$ & $0.00$ / $0.00$ & $0.51$ / $0.49$ & \textemdash & \textemdash & $0.30$ / $0.27$ \\
$p=100$ & $\mathbf{0.02}$ / $\mathbf{0.53}$ & \textemdash & $0.75$ / $0.14$ & $0.00$ / $0.00$ & $0.54$ / $0.37$ & \textemdash & \textemdash & $0.41$ / $0.26$ \\
$\lvert\mathrm{pa}\rvert=0$ & $0.01$ / \textemdash & $0.04$ / \textemdash & $0.99$ / \textemdash & $0.00$ / \textemdash & $0.74$ / \textemdash & \textemdash & \textemdash & $0.16$ / \textemdash \\
$\lvert\mathrm{pa}\rvert=6$ & $\mathbf{0.03}$ / $\mathbf{0.26}$ & $0.31$ / $0.22$ & $0.75$ / $0.06$ & $0.00$ / $0.00$ & $0.52$ / $0.25$ & \textemdash & \textemdash & $0.38$ / $0.15$ \\
$\lvert\mathrm{ch}\rvert=0$ & $\mathbf{0.01}$ / $\mathbf{0.67}$ & $0.16$ / $0.57$ & $0.03$ / $0.40$ & $0.00$ / $0.00$ & $0.15$ / $0.46$ & \textemdash & \textemdash & $0.14$ / $0.41$ \\
$\lvert\mathrm{ch}\rvert=6$ & $\mathbf{0.03}$ / $\mathbf{0.45}$ & $0.46$ / $0.26$ & $0.89$ / $0.07$ & $0.00$ / $0.00$ & $0.69$ / $0.34$ & \textemdash & \textemdash & $0.56$ / $0.19$ \\
Simulated DAG & $0.03$ / $0.46$ & $\mathbf{0.05}$ / $\mathbf{0.83}$ & $0.48$ / $0.21$ & $0.01$ / $0.02$ & $0.27$ / $0.76$ & $0.23$ / $0.36$ & $0.20$ / $0.82$ & $0.13$ / $0.54$ \\
\bottomrule
\end{tabular}

%% file: tables/pc_cells_child.tex
\begin{tabular}{lcccccccc}
\toprule
Setting & PC-PFN & BiDAG & DAGMA & AVICI & PC (Fisher $z$) & PC (RCIT) & GES & MMHC \\
\midrule
Baseline & $\mathbf{0.08}$ / $\mathbf{0.94}$ & $0.26$ / $0.61$ & $0.28$ / $0.23$ & $0.00$ / $0.01$ & $0.17$ / $0.09$ & \textemdash & \textemdash & $0.34$ / $0.48$ \\
$n=100$ & $\mathbf{0.06}$ / $\mathbf{0.69}$ & $0.31$ / $0.47$ & $0.29$ / $0.23$ & $0.01$ / $0.01$ & $0.16$ / $0.07$ & \textemdash & \textemdash & $0.34$ / $0.30$ \\
$n=800$ & $\mathbf{0.07}$ / $\mathbf{0.98}$ & \textemdash & $0.28$ / $0.26$ & $0.00$ / $0.01$ & $0.10$ / $0.08$ & \textemdash & \textemdash & $0.32$ / $0.58$ \\
$p=20$ & $\mathbf{0.07}$ / $\mathbf{0.87}$ & $0.33$ / $0.43$ & $0.39$ / $0.24$ & $0.04$ / $0.13$ & $0.12$ / $0.08$ & \textemdash & \textemdash & $0.38$ / $0.42$ \\
$p=100$ & $\mathbf{0.07}$ / $\mathbf{0.92}$ & \textemdash & $0.17$ / $0.20$ & $0.00$ / $0.00$ & $0.16$ / $0.12$ & \textemdash & \textemdash & $0.31$ / $0.47$ \\
$\lvert\mathrm{pa}\rvert=0$ & $\mathbf{0.03}$ / $\mathbf{0.98}$ & $0.05$ / $0.77$ & $0.00$ / $0.14$ & $0.00$ / $0.01$ & $0.00$ / $0.19$ & \textemdash & \textemdash & $0.01$ / $0.75$ \\
$\lvert\mathrm{pa}\rvert=6$ & $\mathbf{0.10}$ / $\mathbf{0.91}$ & $0.32$ / $0.62$ & $0.21$ / $0.31$ & $0.00$ / $0.00$ & $0.12$ / $0.08$ & \textemdash & \textemdash & $0.33$ / $0.46$ \\
$\lvert\mathrm{ch}\rvert=0$ & $0.04$ / \textemdash & $0.27$ / \textemdash & $0.12$ / \textemdash & $0.00$ / \textemdash & $0.41$ / \textemdash & \textemdash & \textemdash & $0.61$ / \textemdash \\
$\lvert\mathrm{ch}\rvert=6$ & $\mathbf{0.08}$ / $\mathbf{0.93}$ & $0.29$ / $0.62$ & $0.19$ / $0.26$ & $0.00$ / $0.01$ & $0.14$ / $0.07$ & \textemdash & \textemdash & $0.22$ / $0.34$ \\
Simulated DAG & $0.04$ / $0.74$ & $\mathbf{0.09}$ / $\mathbf{0.95}$ & $0.46$ / $0.58$ & $0.00$ / $0.02$ & $0.04$ / $0.45$ & $0.10$ / $0.25$ & $0.21$ / $0.94$ & $0.21$ / $0.79$ \\
\bottomrule
\end{tabular}

%% file: tables/rd_biomarkers.tex
\begin{tabular}{lccccccccc}
\toprule
Task & $n$ & $p$ & CD-PFN & Spike-and-slab & SuSiE & IPSS & DART & Lasso & Correlation \\
\midrule
ER & $1043$ & $456$ & $\mathbf{1}$ & $\mathbf{1}$ & $\mathbf{1}$ & $\mathbf{1}$ & $1$ & $1$ & $1$ \\
PR & $1040$ & $456$ & $\mathbf{1}$ & $\mathbf{2}$ & $\mathbf{1}$ & $\mathbf{2}$ & $2$ & $1.5$ & $1$ \\
HER2 & $893$ & $456$ & $\mathbf{1}$ & $\mathbf{2}$ & $\mathbf{1}$ & $\mathbf{1}$ & $1$ & $1$ & $1$ \\
Lobular & $985$ & $456$ & $\mathbf{1}$ & $\mathbf{1}$ & $\mathbf{1}$ & $\mathbf{1}$ & $2$ & $1$ & $1$ \\
MSI stomach & $383$ & $456$ & $\mathbf{1}$ & $\mathbf{1}$ & $\mathbf{1}$ & $\mathbf{1}$ & $2$ & $1$ & $1$ \\
MSI colorectal & $344$ & $456$ & $\mathbf{1}$ & $\mathbf{1}$ & $\mathbf{1}$ & $\mathbf{1}$ & $1.5$ & $1$ & $1$ \\
MSI endometrial & $167$ & $456$ & $\mathbf{1}$ & $\mathbf{1}$ & $\mathbf{1}$ & $\mathbf{1}$ & $1$ & $1$ & $1$ \\
HPV & $487$ & $456$ & $\mathbf{2}$ & $10$ & $\mathbf{4}$ & $\mathbf{2}$ & $2$ & $3.5$ & $13$ \\
Triple-negative & $888$ & $456$ & $\mathbf{1}/\mathbf{2}/4$ & $\mathbf{1}/\mathbf{2}/164$ & $\mathbf{1.5}/\mathbf{1.5}/95$ & $\mathbf{1}/\mathbf{4}/235$ & $1.5/1.5/265.5$ & $1.5/4/308.5$ & $1/9/84$ \\
\bottomrule
\end{tabular}

%% file: tables/rd_cellline_single.tex
\begin{tabular}{lccccccccc}
\toprule
Task (truth) & $n$ & $p$ & CD-PFN & Spike-and-slab & SuSiE & IPSS & DART & Lasso & Correlation \\
\midrule
ARID1B (ARID1A) & $1103$ & $668$ & $\mathbf{1}$ & $\mathbf{2}$ & $\mathbf{1}$ & $\mathbf{1}$ & $6.5$ & $1$ & $1$ \\
BRAF (BRAF mut.) & $1103$ & $668$ & $\mathbf{1}$ & $\mathbf{2}$ & $\mathbf{1}$ & $\mathbf{1}$ & $10.5$ & $1$ & $1$ \\
CTNNB1 (CTNNB1 mut.) & $1103$ & $668$ & $15$ & $\mathbf{5}$ & $\mathbf{5}$ & $20$ & $14$ & $9$ & $12$ \\
ERBB2 (ERBB2) & $1103$ & $668$ & $\mathbf{1}$ & $\mathbf{1}$ & $\mathbf{1}$ & $\mathbf{1}$ & $7$ & $1$ & $1$ \\
ESR1 (ESR1) & $1103$ & $668$ & $\mathbf{1}$ & $\mathbf{2}$ & $\mathbf{1.5}$ & $\mathbf{1}$ & $5.5$ & $1.5$ & $2$ \\
KRAS (KRAS mut.) & $1103$ & $668$ & $\mathbf{1}$ & $\mathbf{1}$ & $\mathbf{1}$ & $\mathbf{1}$ & $11.5$ & $1$ & $1$ \\
PIK3CA (PIK3CA mut.) & $1103$ & $668$ & $\mathbf{1}$ & $\mathbf{4}$ & $\mathbf{1}$ & $\mathbf{1}$ & $9.5$ & $1$ & $1$ \\
SMARCA2 (SMARCA4) & $1103$ & $668$ & $\mathbf{1}$ & $\mathbf{3}$ & $\mathbf{1}$ & $\mathbf{1}$ & $10.5$ & $1$ & $1$ \\
STAG1 (STAG2) & $1103$ & $668$ & $\mathbf{1}$ & $\mathbf{1}$ & $\mathbf{1}$ & $\mathbf{1}$ & $6$ & $1$ & $1$ \\
WRN (MLH1) & $1103$ & $668$ & $89$ & $\mathbf{7}$ & $\mathbf{3}$ & $1$ & $10.5$ & $10$ & $15$ \\
Afatinib (EGFR) & $710$ & $623$ & $2$ & $8$ & $37$ & $\mathbf{2}$ & $4.5$ & $75.5$ & $92$ \\
Alpelisib (PIK3CA mut.) & $710$ & $623$ & $\mathbf{1}$ & $\mathbf{1}$ & $\mathbf{1}$ & $\mathbf{1}$ & $4$ & $1$ & $1$ \\
Dabrafenib (BRAF mut.) & $702$ & $622$ & $\mathbf{1}$ & $\mathbf{2}$ & $\mathbf{1}$ & $\mathbf{1}$ & $18.5$ & $1$ & $1$ \\
Fulvestrant (ESR1) & $708$ & $619$ & $29$ & $46$ & $209$ & $\mathbf{2}$ & $481.5$ & $477.5$ & $9$ \\
Lapatinib (ERBB2) & $710$ & $619$ & $\mathbf{1}$ & $\mathbf{1}$ & $\mathbf{1}$ & $\mathbf{1}$ & $6$ & $1$ & $1$ \\
Nutlin-3a (TP53 mut.) & $712$ & $623$ & $\mathbf{1}$ & $\mathbf{1}$ & $\mathbf{1}$ & $\mathbf{1}$ & $8.5$ & $1$ & $1$ \\
Venetoclax (BCL2) & $704$ & $617$ & $\mathbf{1}$ & $\mathbf{3}$ & $\mathbf{1.5}$ & $\mathbf{1}$ & $28.5$ & $2$ & $2$ \\
\bottomrule
\end{tabular}

%% file: tables/rd_cellline_multi.tex
\begin{tabular}{lccccccccc}
\toprule
Drug & $n$ & $p$ & CD-PFN & Spike-and-slab & SuSiE & IPSS & DART & Lasso & Correlation \\
\midrule
Ipatasertib & $704$ & $617$ & $\mathbf{1}/\mathbf{2}$ & $\mathbf{1}/22$ & $\mathbf{1}/7$ & $\mathbf{1}/17$ & $6.5/6.5$ & $1/13.5$ & $1/22$ \\
MK-2206 & $712$ & $623$ & $2/171$ & $17/401$ & $8/140$ & $5/325.5$ & $13/110$ & $20.5/338.5$ & $127/208$ \\
Nutlin-3a & $712$ & $623$ & $\mathbf{1}/\mathbf{4}$ & $\mathbf{1}/\mathbf{2}$ & $\mathbf{1}/\mathbf{2}$ & $\mathbf{1}/\mathbf{3}$ & $7.5/7.5$ & $1/2$ & $1/2$ \\
PD0325901 & $712$ & $623$ & $\mathbf{1}/\mathbf{2}/\mathbf{4}$ & $\mathbf{2}/\mathbf{3}/\mathbf{4}$ & $\mathbf{1}/\mathbf{2}/\mathbf{4}$ & $\mathbf{1}/\mathbf{3}/\mathbf{12}$ & $7/7/7$ & $1/4.5/14$ & $1/18/54$ \\
Selumetinib & $697$ & $618$ & $\mathbf{1}/\mathbf{3}/\mathbf{4}$ & $\mathbf{1}/\mathbf{3}/20$ & $\mathbf{1}/\mathbf{3}/\mathbf{4}$ & $\mathbf{3}/\mathbf{6}/\mathbf{10}$ & $8/8/16$ & $1/9.5/9.5$ & $1/28/54$ \\
Trametinib & $711$ & $623$ & $\mathbf{1}/\mathbf{3}/\mathbf{6}$ & $\mathbf{3}/14/25$ & $\mathbf{1}/5/6$ & $\mathbf{1}/\mathbf{13}/328$ & $4/4/14$ & $1/9/19.5$ & $1/36/97$ \\
\bottomrule
\end{tabular}

%% file: tables/rd_nhanes.tex
\begin{tabular}{lccccccc}
\toprule
$n$ & CD-PFN & Spike-and-slab & SuSiE & IPSS & BH & Knockoffs & Knockoffs (RF) \\
\midrule
100 & $0.5$ ($0.4$) & $0.1$ ($0.3$) & $0.6$ ($0.8$) & $3.0$ ($2.4$) & $0.0$ ($0.1$) & $0.2$ ($0.3$) & $0.8$ ($0.8$) \\
200 & $1.6$ ($0.8$) & $2.0$ ($3.2$) & $1.4$ ($1.5$) & $4.8$ ($3.9$) & $0.6$ ($1.4$) & $0.5$ ($0.7$) & $0.9$ ($1.1$) \\
500 & $4.0$ ($2.1$) & $2.8$ ($2.1$) & $2.5$ ($2.0$) & $5.2$ ($4.7$) & $1.6$ ($3.0$) & $0.6$ ($0.6$) & $2.8$ ($3.5$) \\
1000 & $5.8$ ($3.0$) & $4.5$ ($3.2$) & $4.3$ ($2.5$) & $6.0$ ($6.0$) & $3.4$ ($2.0$) & $1.4$ ($1.3$) & $2.1$ ($2.3$) \\
\bottomrule
\end{tabular}

%% file: tables/rd_exogenous.tex
\begin{tabular}{lcccc}
\toprule
& \multicolumn{2}{c}{Demographic covariates} & \multicolumn{2}{c}{Demographic responses} \\
\cmidrule(lr){2-3}\cmidrule(lr){4-5}
Method & Parents & Children & Children & Parents \\
\midrule
PC-PFN & 29 & 0 & 16 & 0 \\
BiDAG & \textemdash & \textemdash & 50 & 8 \\
AVICI & 0 & 0 & 0 & 0 \\
DAGMA & 2 & 3 & 2 & 3 \\
MMHC & 11 & 11 & 14 & 8 \\
\bottomrule
\end{tabular}